\documentclass[prl,aps,twocolumn,superscriptaddress,floatfix]{revtex4-2}
\usepackage{amsmath}
\usepackage{amssymb}
\usepackage{amsbsy}
\usepackage{graphicx}
\usepackage{color}
\usepackage{hyperref} 
\usepackage{enumerate}
\usepackage{mathtools}

\def\be{\begin{equation}}
\def\ee{\end{equation}}
\def\bfi{\begin{figure}}      
\def\efi{\end{figure}}
\def\bea{\begin{eqnarray}}
\def\eea{\end{eqnarray}}
\newcommand{\sgn}{\text{sgn}}

\newcommand{\br}{{\bf r}}

\newcommand{\bv}{{\bf v}}
\newcommand{\bn}{{\bf n}}
\newcommand{\bg}{{\bf g}}

\newcommand{\bw}{{\boldsymbol{\omega}}}

\newcommand{\bF}{{\boldsymbol{F}}}

\begin{document}

%\title{Collective drifts of dense granular packings: \\ a (frictional?) ratchet effect induced by structural disorder (defects?)}
\title{Collective drifts in vibrated granular packings: \\ the interplay of friction and structure.}

\author{A. Plati}
\affiliation{Department of Physics, University of Rome Sapienza, P.le Aldo Moro 2, 00185, Rome, Italy}
\affiliation{Institute for Complex Systems - CNR, P.le Aldo Moro 2, 00185, Rome, Italy}

\author{A. Puglisi}
\affiliation{Department of Physics, University of Rome Sapienza, P.le Aldo Moro 2, 00185, Rome, Italy}
\affiliation{Institute for Complex Systems - CNR, P.le Aldo Moro 2, 00185, Rome, Italy}
\affiliation{INFN, University of Rome Tor Vergata, Via della Ricerca Scientifica 1, 00133, Rome, Italy}

\begin{abstract}
We simulate vertically shaken dense granular packings with horizontal periodic boundary conditions. A coordinated translating motion of the whole medium emerges when  the horizontal symmetry is broken by disorder or defects in the packing and the shaking is weak enough to conserve the structure. We argue that such a drift originates in the interplay between structural symmetry breaking and frictional forces transmitted by the vibrating plate.  A non-linear ratchet model with stick-slips reproduces many faces of the phenomenon. The collective motion discussed here underlies phenomena observed recently with vibrofluidized granular materials, such as persistent rotations and anomalous diffusion.
\end{abstract}

\maketitle

\emph{Introduction.}--- One of the major challenges of statistical mechanics, nowadays, is understanding systems far from equilibrium, e.g. in the presence of energy flows or dissipation~\cite{marconi2008fluctuation}. A realm of physics where fluctuations and lack of thermodynamic equilibrium are ubiquitous is soft matter~\cite{nagel2017experimental} and, in this context, an established test-ground for theories
is provided by vibrofluidized granular systems~\cite{Kadanoff99,Jaeger96,RodhesBook,Nagel2017,deArcangelis2019} where energy is continuously injected from an external source and dissipated in friction. When isotropy is broken, a fraction of this energy current can be exploited to realise a ratchet effect~\cite{Reimann2002,Hanggi2009,Costantini2007,eshuis2010experimental,Gnoli2013,Balzan2011} similar to what is seen in active matter~\cite{DiLeonardo2010}. A series of recent granular experiments suggest that spontaneous persistent drifts emerge in the presence of isotropic disorder: this is revealed by superdiffusion~\cite{Scalliet2015}, which has been
connected to transient drifts with very long
relaxation times \cite{Plati2019,Plati2020slow}, or by the
appearance of steady rotations of disks in dense vibrated packings
\cite{Moukarzel2020} and by the emergence of collective motions under
swirling excitations \cite{Zhao2021I,Zhao2021II}. In all these examples 
%friction and spontaneous (but persistent) asymmetries in the packing disorder conspire to convert 
random energy is converted into a steady flow, realising an interesting class of “disordered engines”. 

Here we aim at understanding the general ingredients underlying these effects, focusing on a few simplified
setups and - in the conclusions - on a  model of frictional ratchet.  The mechanism investigated explains the aforementioned phenomena as it represents their translational (instead of rotational)~\cite{Scalliet2015,Plati2019,Plati2020slow}, collective (instead of individual) \cite{Moukarzel2020}  and spontaneous (instead of externally stimulated) \cite{Zhao2021I,Zhao2021II} counterpart.

\begin{figure} 
\centering
\includegraphics[width=0.84\columnwidth,clip=true]{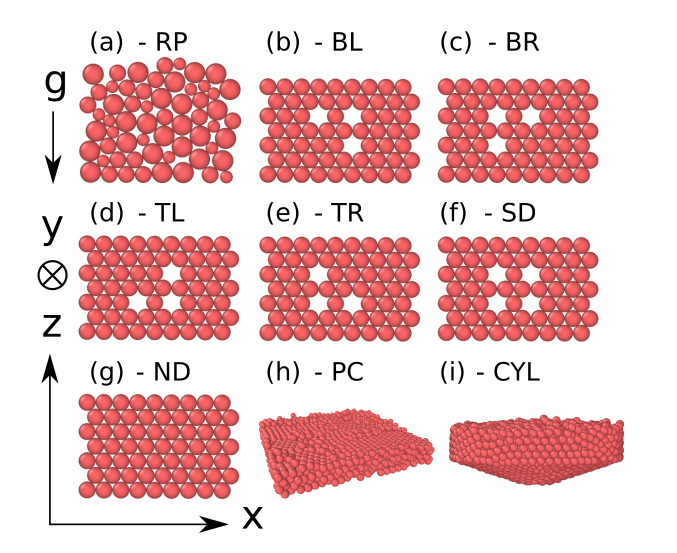}
\caption{Simulation geometries. Quasi 2D vertical layers can be random
  polydisperse (a) -- with equal proportions of three species of
  radii $R_i=\{1.5,2.0,2.5\}$ mm -- or ordered monodisperse (b-g) --
  $R_i=R=2$ mm -- with eventually defects. The defect nomenclature is
  related to the position of the grain that breaks the symmetry. The
  3D setups are cubic (h) and cone-base cylinder (i) (see~\cite{Plati2019} for details),
  both monodisperse. 
  \label{fig:Fig1}}
\end{figure}

\emph{Numerical setup.}--- We simulate, through an established
Discrete Elements Method (DEM), a system of $N$ spherical grains
confined by hard walls and/or periodic boundary conditions (PBC) with
different geometries. Each grain has radius $R_i$, mass $m_i$,
position $r_{\alpha i}$, velocity $v_{\alpha i}$ and angular velocity
$\omega_{\alpha i}$, where $\alpha=\{x,y,z\}$ and $i=\{1,\cdots,N \}$.
The grain-grain and grain-boundary contact forces follow the Hertz
Mindlin (HM) model, see details in the Supplemental Material (SM) \cite{CiteSM} \nocite{Plimpton1995,LammpsSite,Zhang2005,Silbert2001,Brilliantov1996,PopovBook,LammpsSiteGranular,DiMaio2004,PoeschelBook,
Rackl2017,Plati2021Getting,Wilkie2004,Ovito}.  In all cases the setup is enclosed in a 3D box of height $10 \times 2 \max_i R_i$, energy is injected by vertical vibrations $z(t)=A\cos(2\pi f t)$ of the upper and lower confining hard walls, and there is one or more horizontal direction with infinite horizon (i.e. where movement is not constrained by walls). 
In quasi-2D setups (cases in Fig. \ref{fig:Fig1}a-g, $N=60$) $x$ has PBC while $y$ is confined by two parallel vertical walls of width $L=32$ mm  separated by
a distance $d=2\max_i R_i$. In the full 3D case in Fig. \ref{fig:Fig1}h ($N=2600$) both $x$ and $y$ directions have PBC and the base of the box has dimensions $98\times 98$ mm$^2$.
In the full 3D case  in Fig. \ref{fig:Fig1}i ($N=2600$) the box is a cylinder with a conical-shaped
base, identical to what previously used in experiments and simulations
\cite{Scalliet2015,Plati2019,Plati2020slow}.
The shaking intensity is measured by $\Gamma=A(2\pi
f)^2/g$ ,where $g$ is the gravity acceleration. Here we always vary $\Gamma$ by $A$
keeping $f$ constant at 100 Hz in the quasi-2D simulation and at 200 Hz in the all other ones.
 
Regarding the quasi-2D setups, we prepare the initial state of our
packings in two main ways: the first one consists of randomly pouring
a polydisperse assembly of grains in the container, initial velocities
are zero but during the pouring dynamics they acquire energy and
rapidly reach a stationary statistics; the second one is obtained by
placing a monodisperse assembly on an hexagonal lattice with the
possibility to have vacancies in determined sites, while the particles
initial velocities are drawn from a Gaussian distribution with zero
mean and a variance small enough to keep the crystal stable. For the
3D cases we simply pour monodisperse grains in the containers. Since
the latter do not satisfy the right proportions for crystallization
the resulting packings are fairly disordered.  The monodisperse
packings with quasi-2D geometry (b-g in Fig. \ref{fig:Fig1}) can be
without defects (g), or with four symmetric defects (f), or with three
defects placed to break in different ways the symmetry of the crystal
with respect to $z$ and $x$ (b-e). The
polydisperse quasi-2D packing (Fig. \ref{fig:Fig1}a) is inspired by a
recent experimental and numerical study~\cite{Moukarzel2020} with
shaken disks without PBC: for $\Gamma$ low enough
a persistent net angular velocity is seen for each disk. We verified (see SM \cite{CiteSM}) that such
persistent rotational modes are present also in our simulations with
spheres, in all the geometries explored here, with both hard walls and PBC. However, in this Letter, we focus on simulations with horizontal PBC only, where a
different phenomenon, namely the collective horizontal persistent drift, superimposed on the previously observed rotational
modes, appears.

\begin{figure}
\centering
\includegraphics[width=0.48\columnwidth,clip=true]{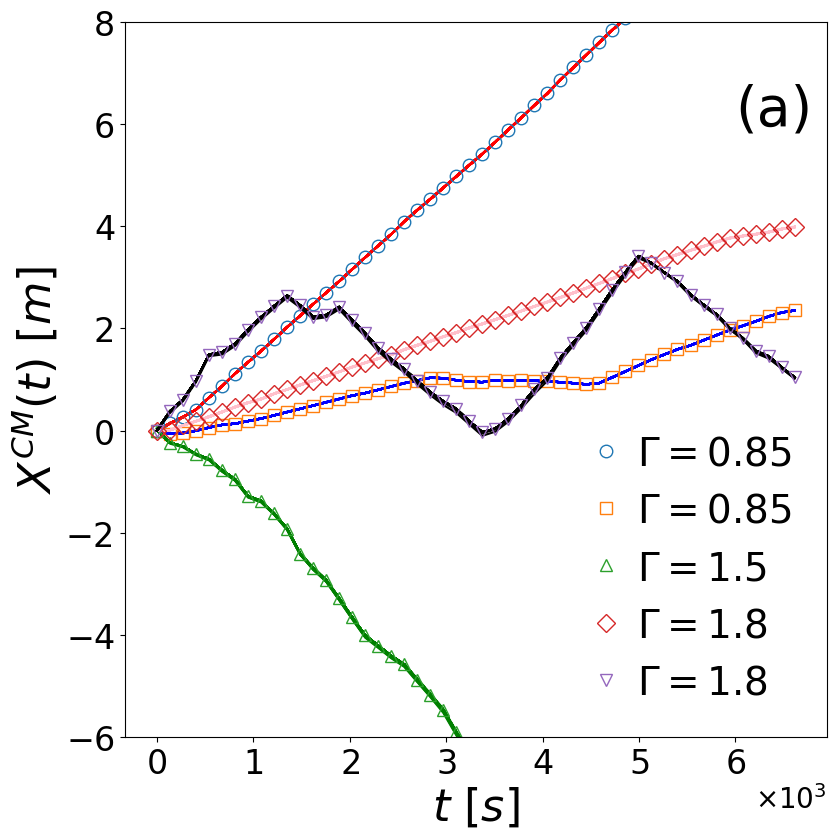}
\includegraphics[width=0.48\columnwidth,clip=true]{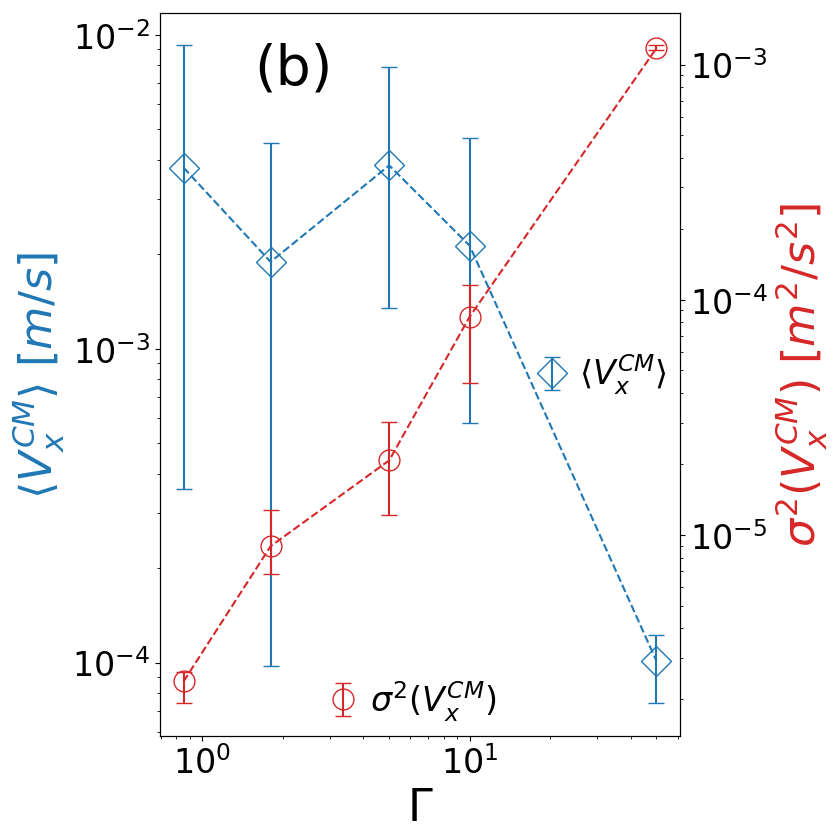}
\includegraphics[width=0.48\columnwidth,clip=true]{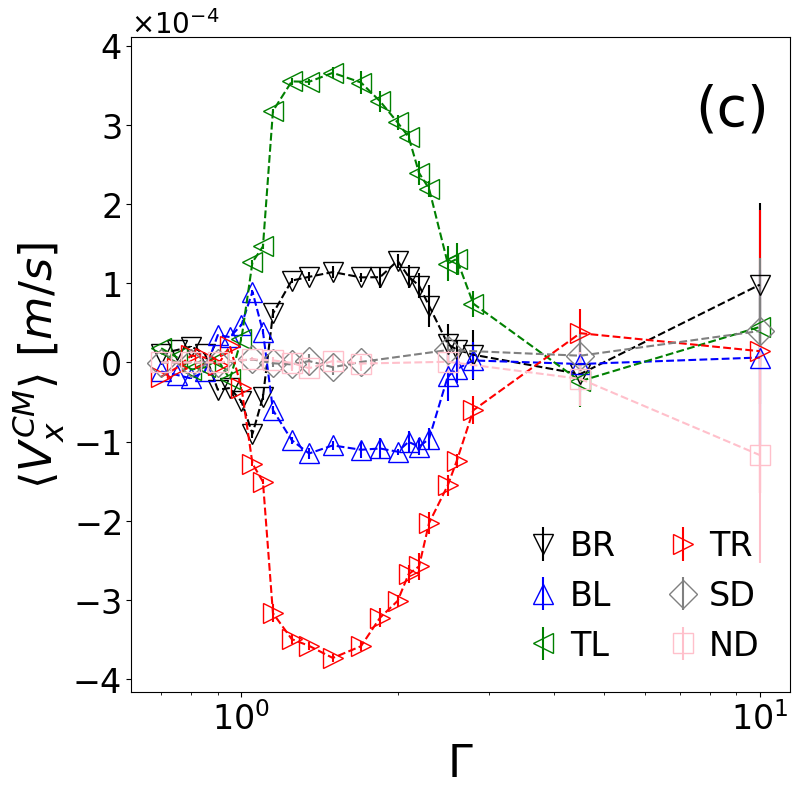}
\includegraphics[width=0.48\columnwidth,clip=true]{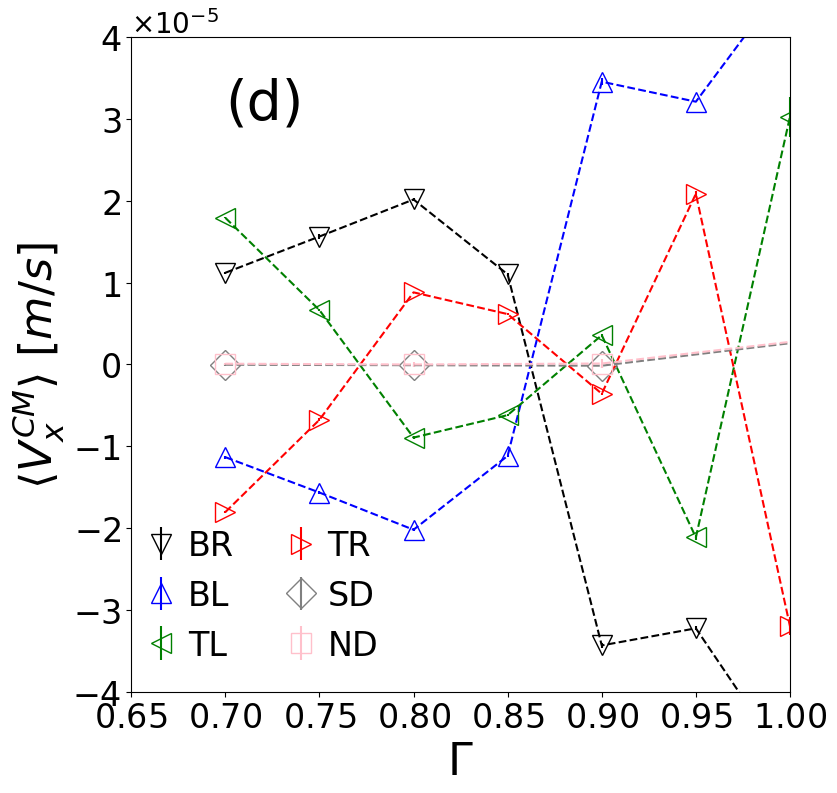}
\caption{a) $X^{CM}(t)$ (symbols) and $x_i(t)$ for 
   all the grains $i$ in the system (lines) for quasi-2D random
  packings: the grains move coherently with the
  CM. Simulations with the same $\Gamma$ refer to different random
  realizations.
% These can lead, with the same driving parameters, to
%  low/high drift velocity (circles and squares) and
%  steady/intermittent motion (diamond and upside-down
%  triangles).
 b) Comparison between $\sigma^2(V_x^{CM})$ and $\langle V_x^{CM} \rangle$, 
  obtained averaging over three independent random realizations.
  c) Time averaged velocity of center of mass as a function of
  $\Gamma$ for different monodispersed packings. Each point is
  mediated over five independent realizations of the dynamics. Here, differently from random
  packings, $\langle V_x^{CM} \rangle$ vanishes for $\Gamma \simeq 3$
  because moderate vibrations destroy the asymmetric configuration of
  defects. d) zoom on low $\Gamma$s to highlight the sign changes
  with fixed structure.\label{fig:Fig2}}
\end{figure}

\emph{Drifting disordered packings - } We start from polydisperse
packings in the quasi-2D geometry. At low values of $\Gamma$ all the
system moves coherently with the center of mass (CM) along the
$x$-axis i.e. the free direction allowed by PBC, whose motion in time
$X^{CM}(t)=M_{\text{tot}}^{-1}\sum_i m_i x_i (t)$ is shown in
Fig. \ref{fig:Fig2}a.  We note that, even with the same $\Gamma$, a
different random packing can lead to a stable drift with very different
magnitudes or to an intermittent drift. In panel b, considering the
$x$ component of the CM's velocity $V_x^{CM}(t)$, we also verify that
short time properties as the time variance
$\sigma^2\left(V_x^{CM}\right)$ are fully determined by the driving
parameters (small errorbars), while the slow cooperative dynamics
(characterized by its time average $\langle V_x^{CM} \rangle$),
sensibly depends on the packing configuration (large
errorbars). Nevertheless, raising $\Gamma$ up to values for which the
system fluidizes (typically $\Gamma > 10$), the CM performs
Brownian-like trajectories and $\langle V_x^{CM} \rangle$ vanishes (SM-Video1 \cite{CiteSM}). We recall that a dense
vibrofluidized granular system has several time-scales, the
smallest associated to fast vibrational motion, the largest associated
to slow rearrangements of the global contact network. What we observe here is a rapid divergence of the largest
time-scale when $\Gamma$ is reduced below $\sim 3 \div 10$. The value of $\langle V_x^{CM} \rangle$, therefore, does not depend crucially on
the trajectory's duration, provided it is longer than the small
time-scales, e.g. $\gg 10^{-1}$ s.
%In the conclusions we discuss of the connection of
%the results presented here with previous (experimental, numerical and
%theoretical) results on anomalous diffusion in vibrofluidised granular
%systems \cite{Scalliet2015,Plati2019,Plati2020slow,Plati2021LongRange}
%is postponed to the conclusions.

\emph{Effect of symmetries - } The observation that average motion is
erased by fast particle rearrangements (fluidization) is a  hint
of its correlation with the system's spatial configuration. For this
reason we study $\langle V_x^{CM} \rangle$ for ordered packings with defects (Fig. \ref{fig:Fig2}c).
Net drifts are never observed in the FC and SD cases, whereas for all asymmetric configurations they are.
Moreover, when the defect configuration is mirrored with
respect to the $z$-axis, e.g. BL $\to$ BR or TR $\to$ TL
etc., at a given value of $\Gamma$, $\langle V_x^{CM} \rangle$
changes its sign remaining with comparable magnitude. A reflection of
the configuration with respect to $x$ always changes the magnitude of
the drift but the sign reverses just for a few values of
$\Gamma$. Remarkably, a variation of $\Gamma$ with keeping the same
layout of defects brings to multiple inversions of $\sgn(\langle
V_x^{CM} \rangle)$ in the region $0.8 \leq \Gamma \leq 1.05$
zoomed in Fig. \ref{fig:Fig2}d. In summary: i) defect
asymmetry is needed to observe non-zero mean velocity
of the CM, ii) reflection w.r.t. $z$ inverts the direction of motion,
iii) the defect configuration alone does not define the verse of the
motion, the driving parameter determines it too (SM-Video2 \cite{CiteSM}).
\begin{figure*}
\centering
\includegraphics[width=0.245\textwidth,clip=true]{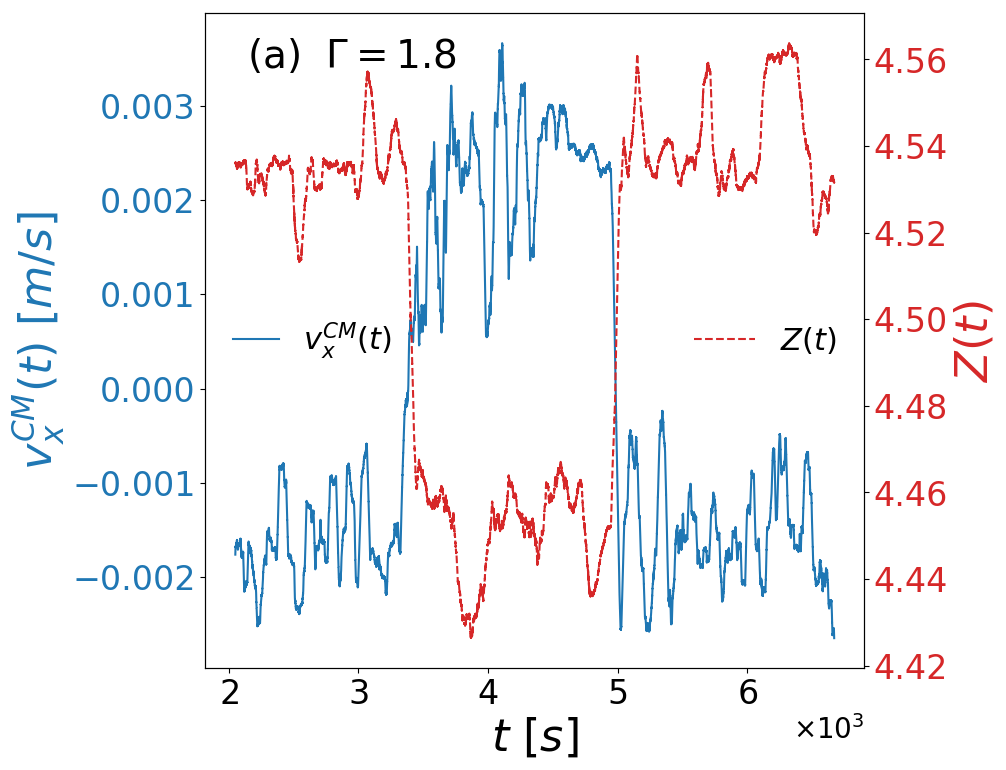}
\includegraphics[width=0.245\textwidth,clip=true]{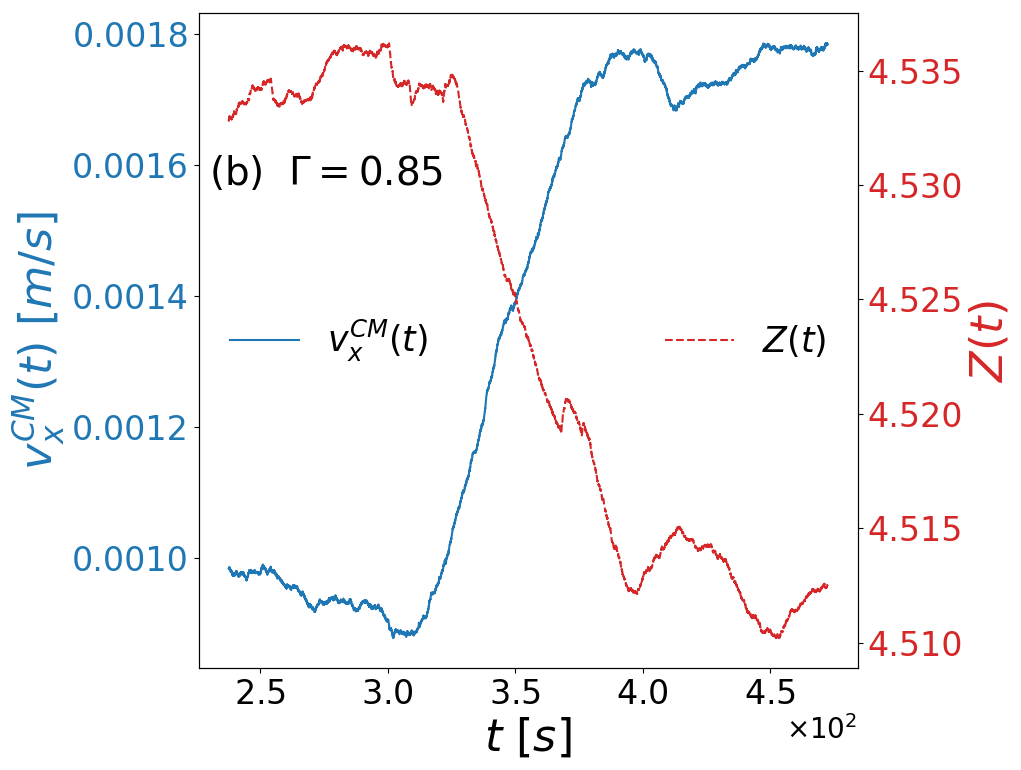}
\includegraphics[width=0.23\textwidth,clip=true]{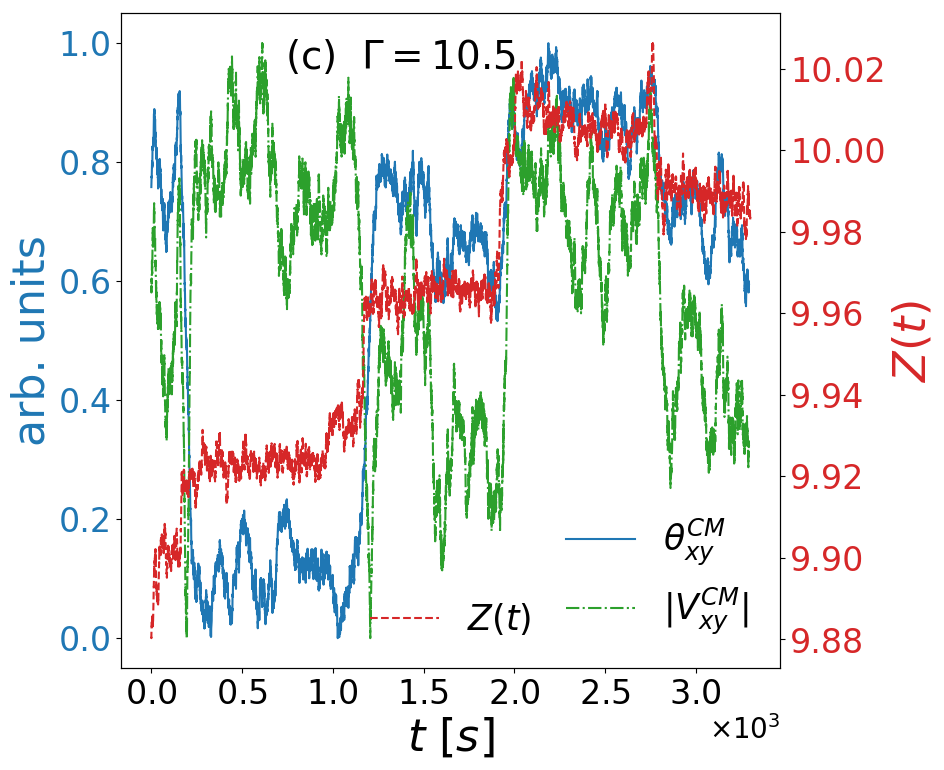}
\includegraphics[width=0.245\textwidth,clip=true]{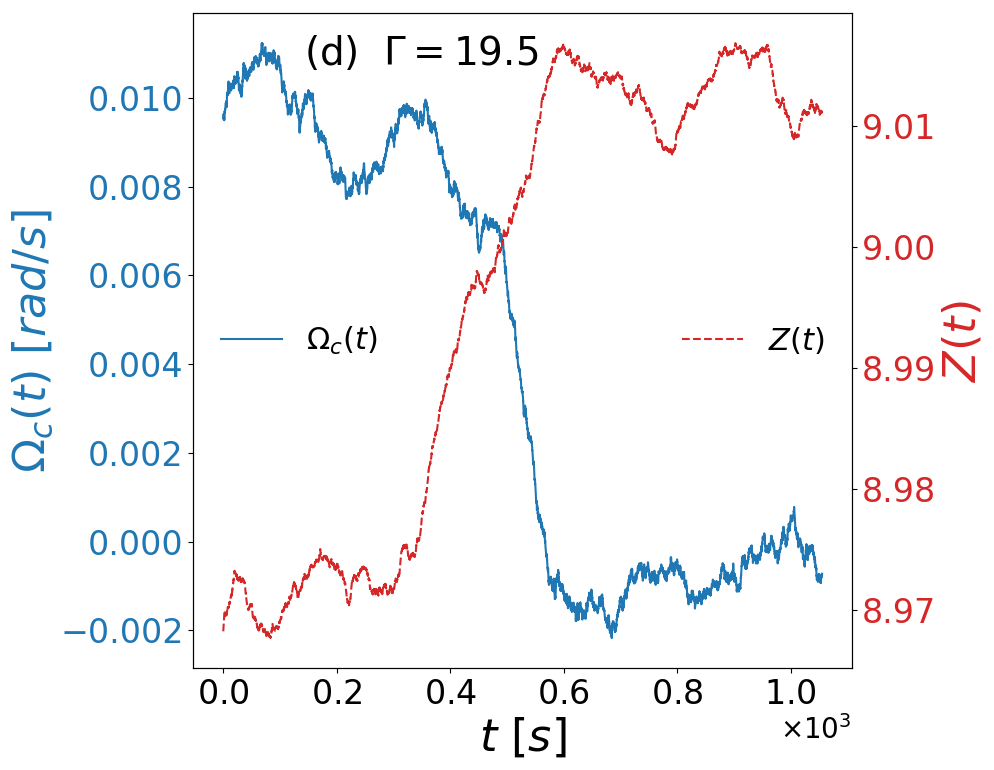}
 \includegraphics[width=0.265\textwidth,clip=true]{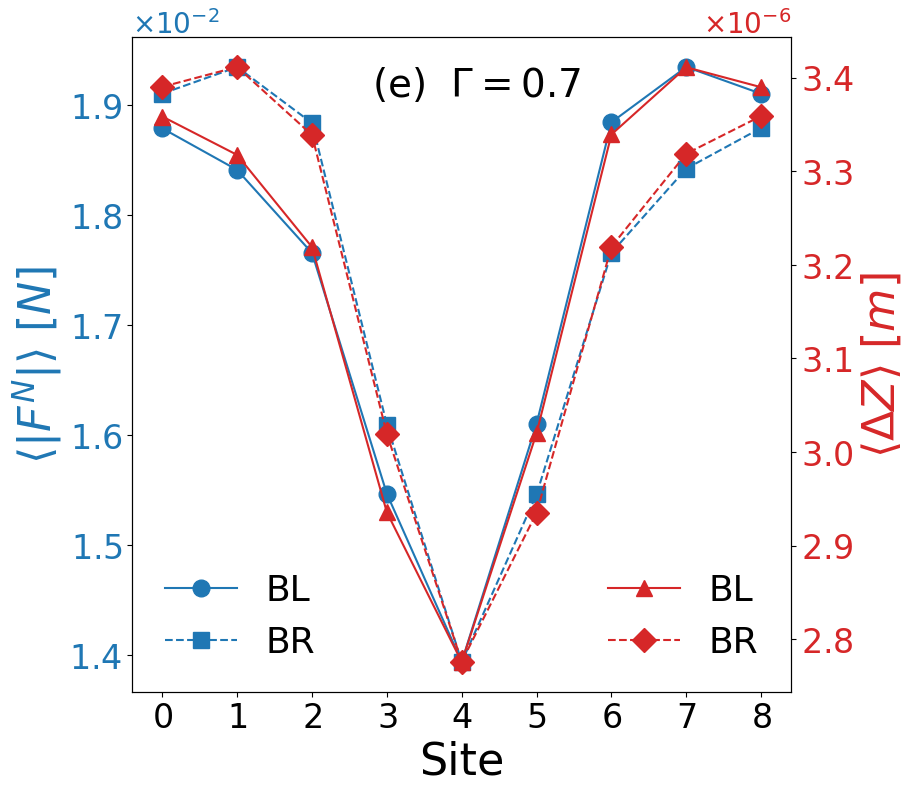}
\includegraphics[width=0.225\textwidth,clip=true]{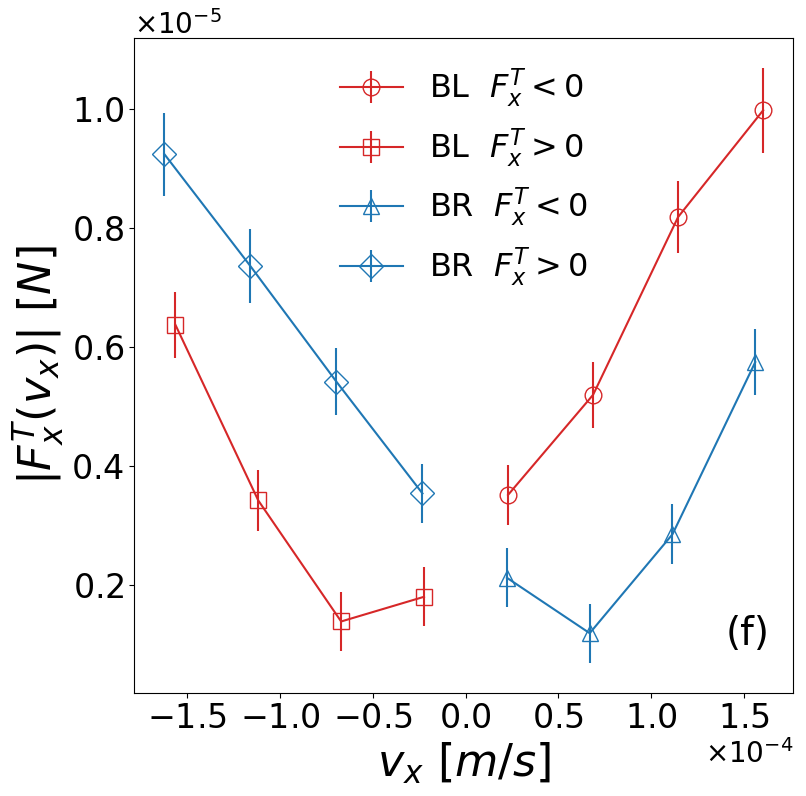}
\includegraphics[width=0.235\textwidth,clip=true]{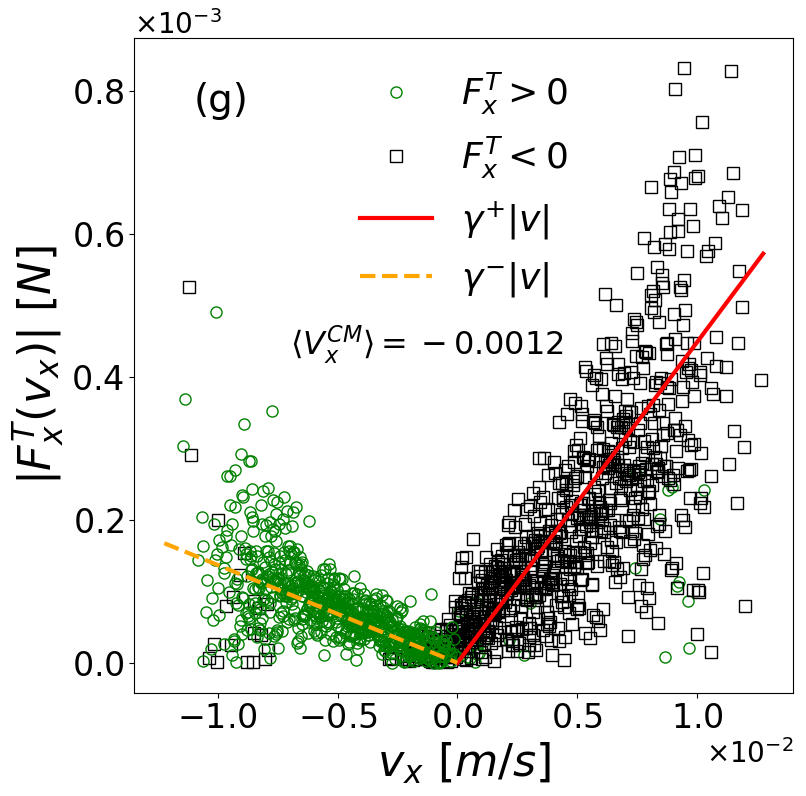}
\includegraphics[width=0.235\textwidth,clip=true]{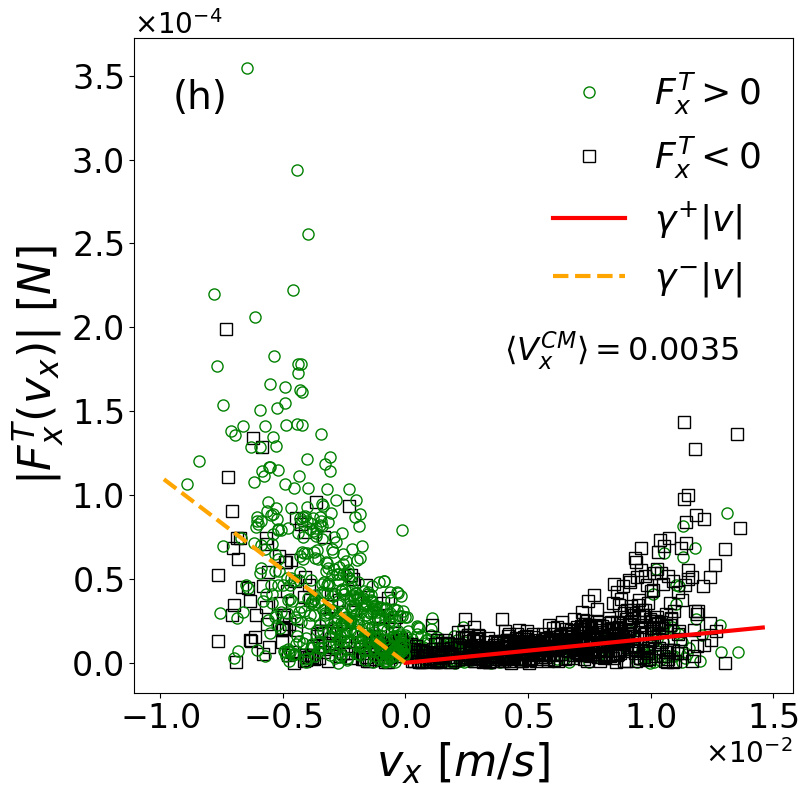}
\caption{a-d) Comparison between the time evolution of the CM velocity
  and the mean coordination number of the contact network
  $Z(t)=N^{-1}\sum_{ij}\Theta(R_{i}+R_{j}+\delta-|\boldsymbol{r}_{ij}|)$
  where $0<\delta \ll R_i+R_j$ allows to detect nearest neighbour not
  in contact due to vibrations. Trajectories are smoothed with a
  running average. a-b) random packings (cases with upside-down triangles and
  circles in Fig. \ref{fig:Fig2}a). c) 3D cubic setup, here
  $V_{xy}^{CM}$ is two dimensional so we plot both the modulus and the
  orientation $\theta_{xy}^{CM}$. d) 3D cylindrical geometry,
  the collective motion is a rotation around the central axis so
  $\Omega_c=N^{-1}\sum_i
  |\boldsymbol{r}_i|^{-2}(\boldsymbol{r}_i\times\boldsymbol{v}_i)_z$.
  e) Time averaged modulus of the normal force and compenetration
  between the plate and the bottom particles for two $z$-reflected
  configurations of defects. Site 4 is under the double defect (see Fig
  \ref{fig:Fig1}b-c). f-h) Scatter plots with bins of the total
  $x$-tangential force modulus exerted by the plate VS the total
  velocity of the bottom particles for ordered packings with
  defects (f) and random packings (g-h) shaken at $\Gamma=0.7$. In the
  latter, $\gamma^{\pm}=|\langle F^T_x | v_x\gtrless
  0\rangle|/\langle v_x | v_x\gtrless
  0\rangle$   linearly interpolate the data clouds.
%conditioned by $g^T_x\gtrless 0$
  \label{fig:Fig3}}
\end{figure*}

\emph{Sensitivity to small configuration changes - } Given the
observed complicate entanglement of external parameters and drifts, we
now focus on fluctuations of the drift during the same
trajectory. Such fluctuations are more evident and frequent in random
packings: Fig. \ref{fig:Fig3}a-b compares the coordination number of
the contact network $Z(t)$ with the instantaneous drift $V_x^{CM}(t)$:
they often change together. Such changes do not correspond to a
significant rearrangement of particles (changes of $Z(t)$ can be
smaller than $1\%$), confirming that the collective motion is sensible
to small deformations of the contact network. Collective drifts are observed also in the 3D realistic setups (see \cite{Plati2019,Plati2020slow} and SM \cite{CiteSM}) and similar correlations are present here too (Fig. \ref{fig:Fig3}c-d). Obviously, the coordination number is not
directly related to the packing asymmetry. A more appropriate quantity eluded our analysis, perhaps because of the observed challenging scenario: on one hand, the drift may
significantly change at fixed $\Gamma$ for weak deformations of the
contact network (as seen for random packings) and, on the other hand,
it can change direction by varying the driving parameters with a fixed
structure (as occurs in the ordered packings with defects).  Analogous
difficulties  have been
discussed recently for glassy systems~\cite{Diaz2021,Coulais2014,Tong2018,Cubuk2015,Hentschel2019,Schwartzman2019}. In our study,
the athermal nature of the system (that needs a mechanical driving to
reach a stationary state) is a peculiar feature that has no
counterpart in thermal glasses~\cite{gradenigo2010ratchet}.

\emph{Bulk asymmetries originate a ratchet effect - } We now look for observables that mediate between bulk
structure and dynamics. The profiles, along $x$ of the mean normal
pressure and the mean compenetration between the plate and the bottom
particles are shown in Fig. \ref{fig:Fig3}e, for an ordered packing
with defects. Each profile is asymmetric and properly inverted under
a $z$-reflection: defects in the bulk actually influence the way in
which the boundaries interact with the source of energy. This may be
expected, since both normal pressure and compenetration affect the
tangential component of the force exerted by the vibrating plate (see
SM \cite{CiteSM}): what is remarkable in our opinion is its global dynamical
consequence. Recent analytical results~\cite{Plati2021LongRange} for
an idealized model of vibrofluidized dense granular material suggest
long-range spatial correlations that may justify the global influence
of defects. A deeper insight is provided by
the external tangential force provided by the plate $F^T_x$ ($x$-component), which directly affects the CM: 
$\dot{V}^{\text{CM}}_x=M_{\text{tot}}^{-1}F^{T}_x(r_{xj},r_{zj},v_{xj},\omega_{yj})$ where the index $j$ refers to the bottom particles (for ordered packings $j\in [1,9]$). 
%$\dot{V}^{\text{CM}}_x=M_{\text{tot}}^{-1}F^{T}_x(r_{x(1-b)},r_{z(1-b)},g^T_{x(1-b)})$
%$\dot{V}^{\text{CM}}_x=M_{\text{tot}}^{-1}F^{T}_x(r_{x1},r_{z1},\cdots,r_{xb},r_{zb},g^T_{x1}\cdots,g^T_{xb})$
%\begin{equation}\label{eq:VCM}
%\dot{V}^{\text{CM}}_x=M_{\text{tot}}^{-1}F^{T}_x(r_{x1},r_{z1},\cdots,r_{xb},r_{zb},g^T_{x1}\cdots,g^T_{xb}),
%\end{equation}
We characterise the dependence of $F^T_x$
upon $v_{x}=\sum_{j}v_{xj}$ by the scatter plot  in
Fig. \ref{fig:Fig3}f for two $z$-reflected ordered configurations with
defects. Such
dependence is typical of a frictional force but, quite surprisingly,
its intensity depends on the sign of the velocity. Moreover, also this
frictional asymmetry is inverted under $z$-reflection of packing.  
Comparing with Fig. \ref{fig:Fig2}d clarifies that the drift occurs
towards lower friction. Panels g-h show that the same kind of bias is more pronounced for random packings, indeed $|\langle V_x^{CM} \rangle|$ is far larger. All
this makes clear that structural disorder and defects introduce an
asymmetric interaction with the plate. The role of the $\omega_{yj}$s is discussed in the SM \cite{CiteSM}.
\begin{figure}
\centering
\includegraphics[width=0.44\columnwidth,clip=true]{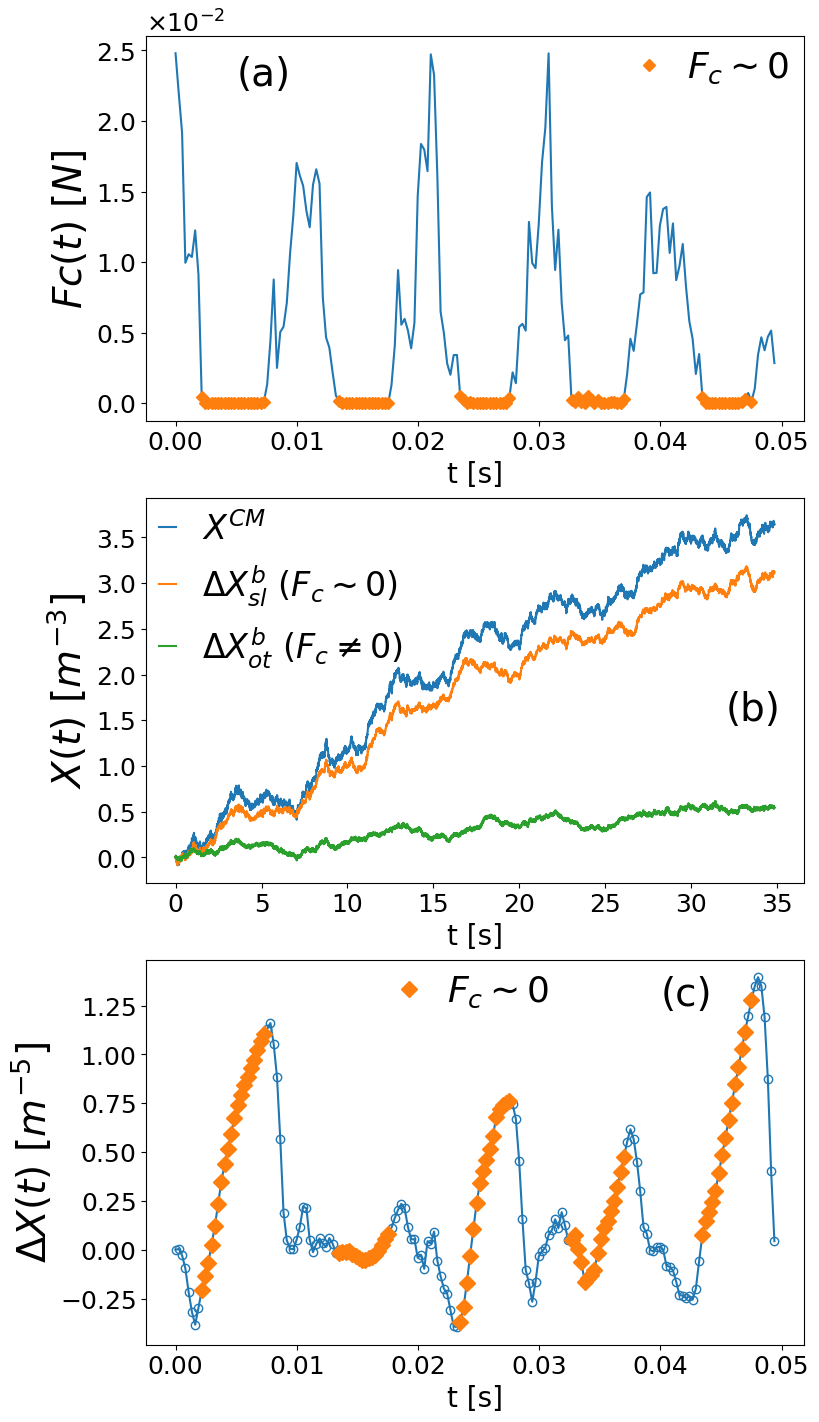}
\includegraphics[width=0.53\columnwidth,clip=true]{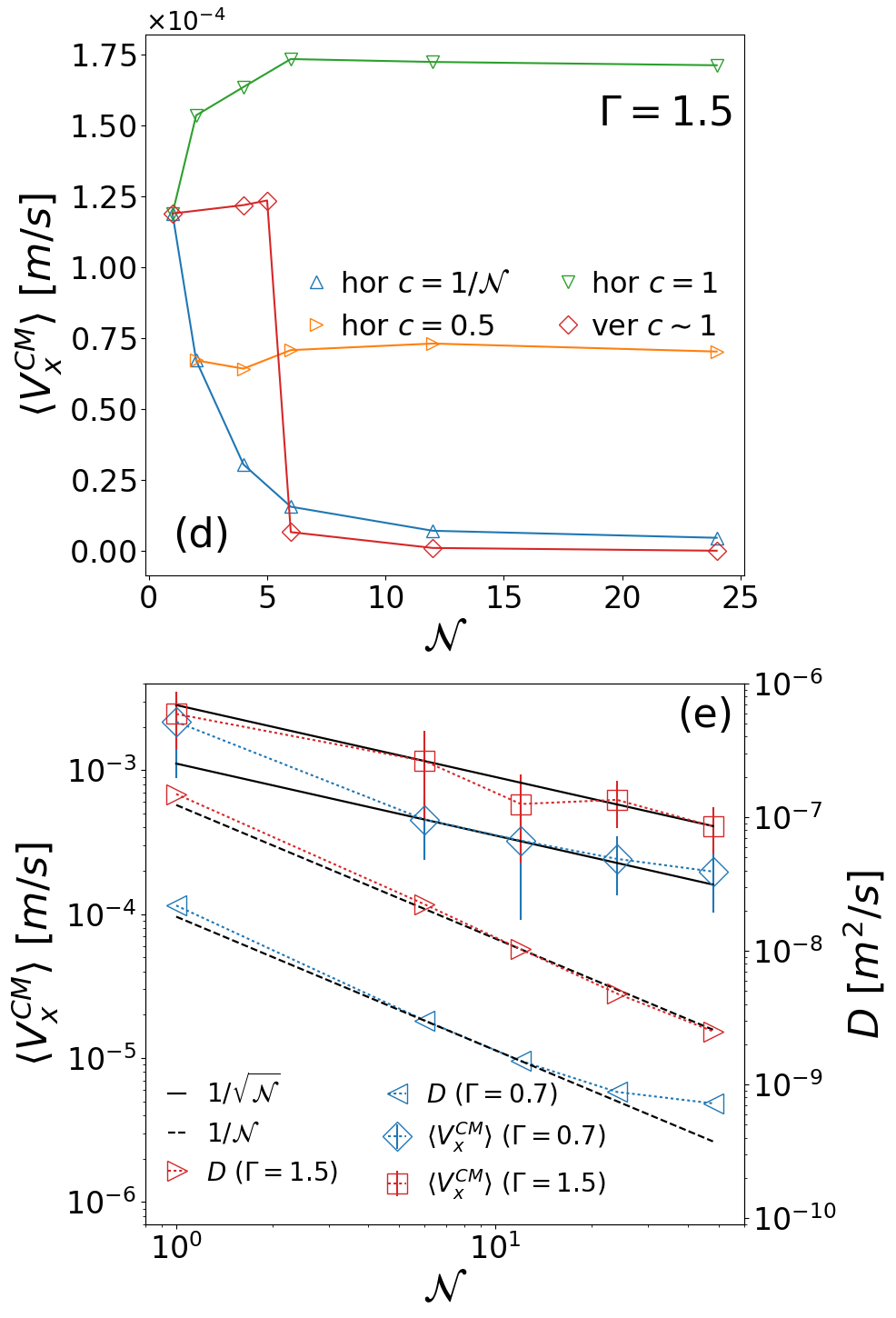}
\caption{a) Time evolution of $F_c=\mu|F^{N}| - |F^{T}_x|$. Diamonds mark the instants for which $F_c \sim 0$ corresponding to the sliding condition for the tangential force between the plate and the bottom grains. b) Trajectory of the $X^{CM}$ compared with the horizontal displacement of the bottom layer accumulated in the sliding instants ($\Delta X^b_{\text{sl}}$) and in all the other ones ($\Delta X^b_{\text{ot}}$). c) Stick-slip in the difference of the mean $x$-coordinates of the two lowest horizontal layers in a granular simulation. Diamonds refer to sliding instants.  These three panels are obtained with BL and $\Gamma=1.25$. d) Mean velocity of the CM as a function of the horizontal (triangles) and vertical (diamonds) size of the system for different values of the defect concentration. The vertical size effect is shown for $c=1$ but we verified that its relevant behavior does not depend on $c>0$. e) Scaling of $\langle V_x^{CM} \rangle$ and the diffusivity of the fluctuation around the drift $D$ (see SM \cite{CiteSM}) as a function of the horizontal system size for random packings.  Each point is obtained averaging over ten independent realizations of the packing, error bars are standard deviations. 
%METTERE LETTERE IN PANEL - METTERE INDICAZIONI SU GAMMA
\label{fig:Fig4}}
\end{figure}

Simulations with a higher acquisition rate allow to study the grain trajectories on timescales shorter than a driving period $1/f$. In Fig. \ref{fig:Fig4}a, we show $F_c (t)=\mu|F^{N}| - |F^{T}_x|$ where $\mu$ is the dynamic friction coefficient. As explained in the SM \cite{CiteSM}, this difference is zero if the grain and the plate surfaces slide on each other. We observe roughly periodical cycles (with the periodicity of the driving plate), with a part of each period where $F_c \sim 0$. In Fig. \ref{fig:Fig4}b,  we compare the trajectory of the CM with the total horizontal displacement accumulated by the bottom layer considering just the sliding instants and all the remaining ones. From this comparison we conclude that the main contribution to the irreversible motion occurs when $F_c \sim 0$. From Fig. \ref{fig:Fig4}c we also understand how the mean horizontal motion originated at the bottom of the system is transfered to the top: there we plot the difference $\Delta X$ between the mean $x-$coordinate of the lowest layer and of the one just above it, noting a clear stick-slip dynamics. We have verified that, as suggested by the marked points in the graph, this distance slowly increases during the sliding of the bottom layer and then suddenly decreases when the sliding condition ceases.

The provided analysis suggest the interpretation of the phenomenon under study as a ratchet effect originated from structural asymmetries \cite{Eichhorn2009,Kumar2008}. 
%Indeed, we deal with Since we are dealing with a rectified motion originated from internal asymmetries, we find reasonable tointerpret our results as an intrinsic ratchet effect . 
To further support this, 
a variation of the well known periodically rocked ratchet \cite{Hanggi94} is proposed in the SM \cite{CiteSM}. Inspired by the stick-slip dynamics between the layers of the granular packing we have replaced the asymmetric periodic potential with a confining but slipping one. Our model reproduces many features of the phenomenon under study (see SM \cite{CiteSM}).

%Considering that the CM of the system
%can be trapped for small displacements because of local
%deformations of the grains in contact with the plate, we have replaced
%the asymmetric periodic potential with a confining but slipping one. This choice is motivated by the presence of a stick-slip dynamics between the layers of the granular packing in  our simulations, see Fig. \ref{fig:Fig4}a.

\emph{System's size effects-} The results for 2D packings are obtained in a relatively small system in order to establish a direct connection with former experiments \cite{Moukarzel2020}. The presence of analogous behavior also in larger and more realistic 3D systems signals the generality of the phenomenon. Nevertheless, a study of the system's size effects in the simplified geometry is non-trivial since the origin of
the drift and the resistance against it can depend both on bulk
volume (through the concentration of defects) and external
surfaces (through friction and energy input).
%useful to understand the ideal behavior of the observed phenomenon.Futureperspective concern the effects of the system’s size on thedrifts’ magnitude, a problem which 
%Enlarging the size of this system is a non-trivial procedure for two main ponts: i) the asymmetry of the packings depends on the density of the defects, ii) the energy is supplied inhomogeneously so its flux into the system depends on the specific direction in which the system expands. 
%As summarized by Fig. ? 
We performed simulation of
ordered packings with defects where the original "module" of $N \sim 60$ grains is replicated $\mathcal{N}$ times along both the $x$ and the $z$ axis. Each replica can be with or without defects so that a concentration of defects $c$ can be defined as the number of asymmetric modules over $\mathcal{N}$. For the horizontal size-scaling we considered three different values of defect concentration $c=\{1/\mathcal{N},0.5,1\}$. From Fig. \ref{fig:Fig4}d we see that increasing the system size along $x$ with a vanishing concentration of defects ($c=1/\mathcal{N}$) leads to the weakening of the drift. For finite defect concentration instead, we observe a non zero-asymptotic $\langle V_x^{CM} \rangle$ whose magnitude depends on $c$. Regarding the vertical size-scaling, we note that the drift is suppressed for large $\mathcal{N}$. This fact can be reasonably explained considering that, in absence of lateral wall in the $x$-direction, the pressure at the base of the system increases with its height reducing the global mobility. We have accordingly verified that for packings higher than $\mathcal{N}=5$, the sliding condition $\mu|F^{N}| \sim |F^{T}_x|$ is never satisfied.
%raising the height of the system with a fixed bottom area implies that particles at the base feel a larger normal pressure [nota su jannsen effect? NON SO] reducing the mobility of the whole packing. This is also consistent with the presence of tangential Coulomb friction in the contact model.
%We verified that, lowering the total weight of the system using lighter grains, it is possible to observe the drift also at values of $\mathcal{N}$ for which the original system do not translate (see SM). 
The scenario for large random packings is different: the global asymmetry of many disordered granular patches each pushing the whole system in a different direction is expected to decrease with their number. Indeed we observe a reduction of the drift as $1/\sqrt{\mathcal{N}}$ when the system size is scaled horizontally (Fig. \ref{fig:Fig4}e). Interestingly, also the diffusivity of the fluctuations around the average motion decreases with $\mathcal{N}$. We define such a quantity as the diffusion coefficient $D$ of the fast component of the CM dynamics that can be measured from the power spectral density of $V^{CM}_x(t)$ as shown in the SM \cite{CiteSM}. For $\Gamma > 1$ it follows a scaling $D \sim 1/\mathcal{N}$ while for $\Gamma < 1$ it seems to reach a minimum at large $\mathcal{N}$. In the former case the quantity $D/\langle V_x^{CM} \rangle^2$ that is is the typical time after which the drift (even very small) becomes visible is independent from $\mathcal{N}$. This observation confirms that the drift  is more relevant than a finite-size effect and poses an interesting challenge for the recently discovered Thermodynamic Uncertainty Relations~\cite{barato2015thermodynamic,horowitz2020thermodynamic}. 

\emph{Conclusion.}--- We have presented a numerical study of a
realistic granular contact model with several ideal experiments to put
in evidence the existence of a random-to-direct energy conversion
based upon the concurrent breaking of time and space symmetries. As
in many ratchet-like phenomena, the same geometry may lead to opposite
drifts, depending on energy injection. Inspired by the interplay of friction and structure emerged in the numerical analysis, we introduce a novel ratchet model with asymmetric interactions and stick-slips. The drifts discovered and
explained underlie several phenomena observed recently in shaken granular media~\cite{Scalliet2015,Plati2019,Plati2020slow,Moukarzel2020} and are expected to be a general feature of soft matter systems, such as active matter, crowd dynamics etc.
%Future perspective concern the effects of the system’s size on the
%drifts’ magnitude, a problem which is non-trivial since the origin of
%the drift and the resistance against it both can depend both on bulk
%volume (through the concentration of defects) and external
%surfaces (through friction). 
%Also the study of Thermodynamic
%Uncertainty Relations (TUR) related to these drifts is a promising
%direction~\cite{barato2015thermodynamic,horowitz2020thermodynamic}.\\

\acknowledgments{The authors are indebted to Marco Baldovin and Rafael Díaz Hernández Rojas for fruitful scientific discussions. The authors acknowledge the financial support from the MIUR PRIN2017 project 201798CZLJ}

\bibliographystyle{apsrev4-2}
\bibliography{BiblioGranRatch}

\clearpage
\onecolumngrid

{\bf SM: SUPPLEMENTAL MATERIAL}

%\vspace{1cm}
\section{Model for contact dynamics}

We develop our DEM simulations through the LAMMPS package \cite{Plimpton1995,LammpsSite} using the Hertz-Mindlin model
\cite{Zhang2005,Silbert2001,Brillantov1996} to solve the
dynamics during the particle-particle and particle-container contact. This visco-elastic
model takes into account both the elastic and the dissipative response
to the mutual compression between the grains. In addition, it includes
in the dynamics not only the relative translational motion but also the
rotational one. The contact forces are described by Eqs. \eqref{eq::CAP2::ForzaHM} below.
\begin{subequations}\label{eq::CAP2::ForzaHM}
	\begin{align}	
    \bF^{N}_{ij}= \sqrt{R_{ij}^{\text{eff}}}\sqrt{\xi_{ij}(t)}\left[(k_{n}\xi_{ij}(t)-m^{\text{eff}}_{ij}\gamma_{n}\dot{\xi}_{ij}(t))\cdot
\bn_{ij} (t)\right] \\
    \bF^{T}_{ij}=\begin{cases}
    \bF^{\text{hist}}_{ij} \quad\text{if}\quad |\bF^{\text{hist}}_{ij} | \le |\mu\bF^{N}_{ij}|     \\
    -\dfrac{|\mu\bF^{N}_{ij}|}{|\bg^{T}_{ij}(t)|}\cdot \bg^{T}_{ij}(t) \quad \text{otherwise} \\
	\end{cases}    
	\\
\bF^{\text{hist}}_{ij}=-\sqrt{R_{ij}^{\text{eff}}}\left[k_t \displaystyle\int \sqrt{\xi_{ij}(t')}\textbf{ds}_{ij}(t')+m^{\text{eff}}_{ij}\gamma_{t}\sqrt{\xi_{ij}(t)}\bg^{T}_{ij}(t)\right]
	\end{align} 
 \end{subequations}

These equations are written for two spherical particles with radius
$R_i$, $R_j$, mass $m_i$,$m_j$ position $\br_i$, $\br_j$, translational
velocity $\bv_i$, $\bv_j$ and rotational velocity
$\bw_i$, $\bw_j$.  The relative velocity is defined
as $\bg_{ij}= (\dot{\br}_{i}-\bw_{i}\times
R_{i}\bn)-(\dot{\br}_{j}+\bw_{j}\times R_{j}\bn)$
where
$\bn=\left(\br_{i}-\br_{j}\right)/\left|\br_{i}-\br_{j}\right|$ defines the normal unitary vector of the contact surface;
we call $\bg_{ij}^{N}$ and $\bg_{ij}^{T}$ the two components of $\bg_{ij}$ respectively normal and parallel to $\bn$.  The
instantaneous overlap between the two grains is represented by
$\xi_{ij}(t)=R_{i}+R_{j}-|\br_i-\br_{j}|$ and its derivative
is $\dot{\xi}_{ij}(t)=|\bg^{N}_{ij}|$. Regarding the effective parameters we have $R_{ij}^{\text{eff}}=R_iR_j/(R_i+R_j)$ and $m_{ij}^{\text{eff}}=m_i,m_j/(m_i+m_j)$; here is important to mention that the wall surfaces are treated as spheres with infinite mass and radius. During the contact, the
particles are subjected to a normal force $\bF^{N}_{ij}$ and a
tangential one $\bF^{T}_{ij}$; both these components have an
elastic and a dissipative contribution multiplied by
the couples of parameters $k_{n}$, $k_{t}$ and $\gamma_{n}$, $\gamma_{t}$, respectively.  In the
normal force $\bF^{N}_{ij}$ we can see an elastic term that
descends from the Hertzian theory of contact mechanics
\cite{PopovBook} characterized by a non-linear dependence on the
displacement.
The elastic part of the tangential force consists on a memory term that takes into account the past history of the tangential displacement whose infinitesimal contribute is given by $\textbf{ds}_{ij}(t)$. 
Regarding the cases in Eq. \ref{eq::CAP2::ForzaHM}b, they implement the Coulomb friction condition with a
coefficient $\mu$. When it is satisfied ($|\bF^{\text{hist}}_{ij} | \le |\mu\bF^{N}_{ij}|$), the typical solid-on-solid sliding force is realized on the contact surface. For more details and possible extensions of the Hertz-Mindlin contact forces implemented in LAMMPS see \cite{LammpsSiteGranular}. 

For the purpose of the main text it is important to note that the tangential force is proportional to the normal overlap $\xi_{ij}$ below the Coulomb threshold and to the normal force modulus $|F_{ij}^N|$ above it. This motivate the analysis of Fig. 4e of the main text.

%[https://docs.lammps.org/pair_granular.html].

\section{Details and calibration of the simulation}

Here we discuss how the parameters of Eqs. \ref{eq::CAP2::ForzaHM} and the simulation time step has been set in relation to the properties of the materials in play. We will provide also some additional details on the numerical setup to complete the description done in the main text. 
For all the simulated geometries we consider steel grains in a container made of plexiglass (PMMA). 
The elastic coefficients for two different materials in contact can be directly derived as 
\begin{equation}\label{eq::CAP2::CostEl}
k_n^{ij}=\frac{4}{3} \left( \frac{1-\nu_i^2}{Y_i}+\frac{1-\nu_j^2}{Y_j} \right)^{-1} \quad k_t^{ij}=8\left(\frac{2-\nu_i}{G_i}+\frac{2-\nu_j}{G_j} \right)^{-1}
\end{equation}
where $Y$ is the Young modulus, $G=\frac{Y}{2(1+\nu)}$ the shear modulus and $\nu$ is the Poisson ratio \cite{Zhang2005,PopovBook,DiMaio2004}. When dealing with just one species we define $k_{n/t}=k^{ii}_{n/t}$. Direct formulas for $\gamma_n$ and $\gamma_t$ are lacking but it is a common strategy to choose them verifying \emph{a posteriori} the good agreement with experimental data \cite{PoeschelBook} . 
A widely accepted criterion to fix the simulation time step is to choose it as a fraction of the Rayleigh time namely the time that a superficial acoustic wave takes to cross a single grain. It is related to the characteristics of the material in this way:

\begin{equation}\label{eq::Cap2::tray}
t_{\text{ray}}=\dfrac{\pi R_{\text{min}} \sqrt{2\left(1+ \nu\right)\rho/Y}}{0.163 \nu +0.8766 },
\end{equation} 
%$dt=0.2t_{\text{ray}}$
where $\rho$ is the grain density and $R_{\text{min}}$ the radius of
the littlest grain in the system \cite{Rackl2017}. With this choice the contact dynamics is properly resolved  because $t_{\text{ray}}$ is
usually much smaller than the typical collision times. In our simulation the simulation time step is $dt=0.2t_{\text{ray}}$

In Tab. \ref{tab:TabParam} we report the numerical values of the parameters where apex $s$ is referred to steel and $p$ to plexiglass. In order to reduce obstructions of the collective drift in the $x$ direction, we removed the tangential interaction between the vertical walls and the grains of the 2D setup by setting the related $k_{t}^{sp}$, $\gamma_{t}^{sp}$ and $\mu^{sp}$ to zero (the values reported in the table refer to the interaction with the bottom plate). It is worth noting that we reduce the Young modulus of steel and PMMA of three and two order of magnitude respectively (their real values are $Y^s\simeq 200 $ Gpa and $Y^p\simeq 3 $ Gpa).
The reason is the need to have a bigger $dt$ in order to decrease the simulation time (note that $t_{\text{ray}} \propto Y^{-1/2}$). 
This is a general strategy in DEM simulation where it is possible to have a comparison with experimental data and verify that the softening of the grains doesn't affect the phenomena under interest. We successfully applied this procedure in the numerical study of the 3D cylindrical system \cite{plati2019,Plati2020slow,Plati2021Getting} where an experimental reference was present so we find reasonable to use this method also in the more ideal setups analyzed in the main text (i.e. the 2D and the cubic 3D  ones). Also the numerical values used for $\gamma_n$ and $\gamma_t$, that cannot be fixed from the material properties, are justified \emph{a posteriori} with this same criterion.

\begin{table}[h]

\centering
\begin{tabular}{|c|c|} 
\hline 
$Y^s$ & 210 Mpa \\ 
\hline 
$\nu^s$ & 0.293 \\ 
\hline 
$\mu^{ss}$ & 0.5\\ 
\hline 
$k_{n}^{ss}$ & 153$\times 10^{6}$ Pa \\ 
\hline 
$k_{t}^{ss}$ & 190$\times 10^{6} $ Pa \\
\hline
$\gamma_{n}^{ss}$ & 3.1$\times 10^{4}$ $(\text{sm})^{-1}$ \\
\hline
$\gamma_{t}^{ss}$ & 9.3$\times 10^{3}$ $(\text{sm})^{-1}$ \\
\hline
$Y^{p}$ & 50 Mpa \\

\hline

\end{tabular} \quad
\begin{tabular}{|c|c|}
\hline
$\nu^{p}$ & 0.37 \\
\hline
$\mu^{sp}$ & 0.5 \\
\hline 
$k_{n}^{sp}$ & 617$\times 10^{5}$ Pa\\ 
\hline 
$k_{t}^{sp}$ & 725$\times 10^{5} $ Pa \\
\hline
$\gamma_{n}^{sp}$ & 1.2$\times 10^{7}$ $(\text{sm})^{-1}$ \\
\hline
$\gamma_{t}^{sp}$ & 1.0$\times 10^{5}$ $(\text{sm})^{-1}$ \\
\hline
$t_{\text{ray}}$ & 6.75$\times 10^{-5}$ s \\
\hline
$dt$ & 1.35$\times 10^{-5}$ s \\
\hline
\end{tabular}

\caption{Numerical values for the material properties and the coefficients of the visco-elastic interaction. We also report the time step $dt$ used for the simulations. This numerical values specifically refers to the 2D and the 3D cubic setups. The vertical walls of the 2D setup have $k_{t}^{sp},\gamma_{t}^{sp},\mu^{sp}=0$ so the don't exert any tangential force on the grains. This is the reason why in the Newton equation of the CM reported in the main text only the bottom particles are considered. For the cylindrical system we refer to the Supplemental Materials of \cite{plati2019}.}
\label{tab:TabParam}

\end{table}

\begin{figure}[h!]
\centering
\includegraphics[width=0.35\textwidth]{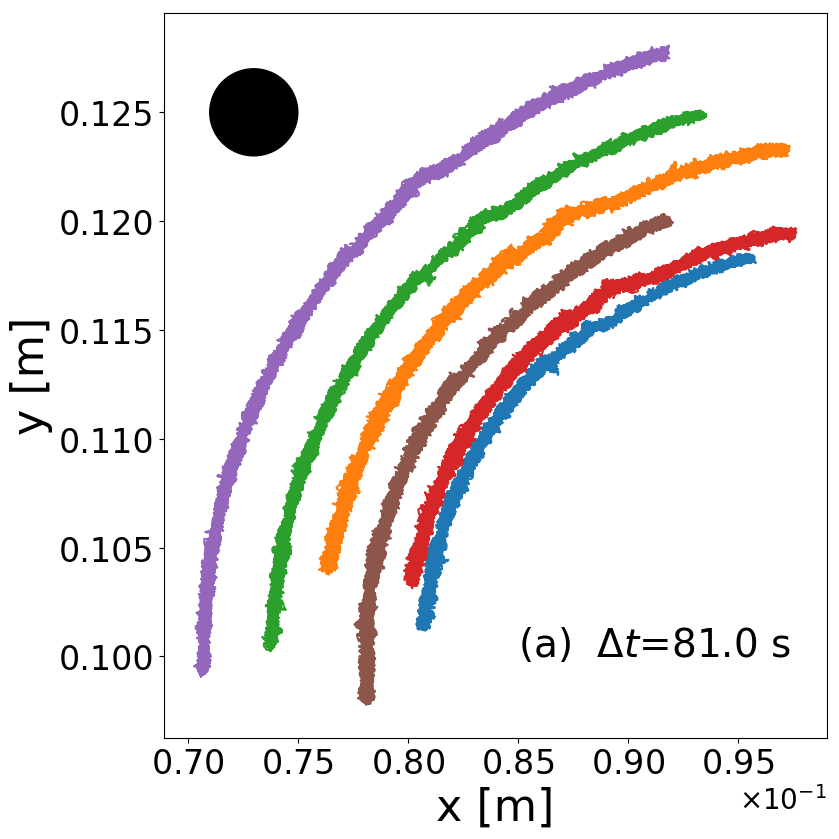}
\includegraphics[width=0.35\textwidth]{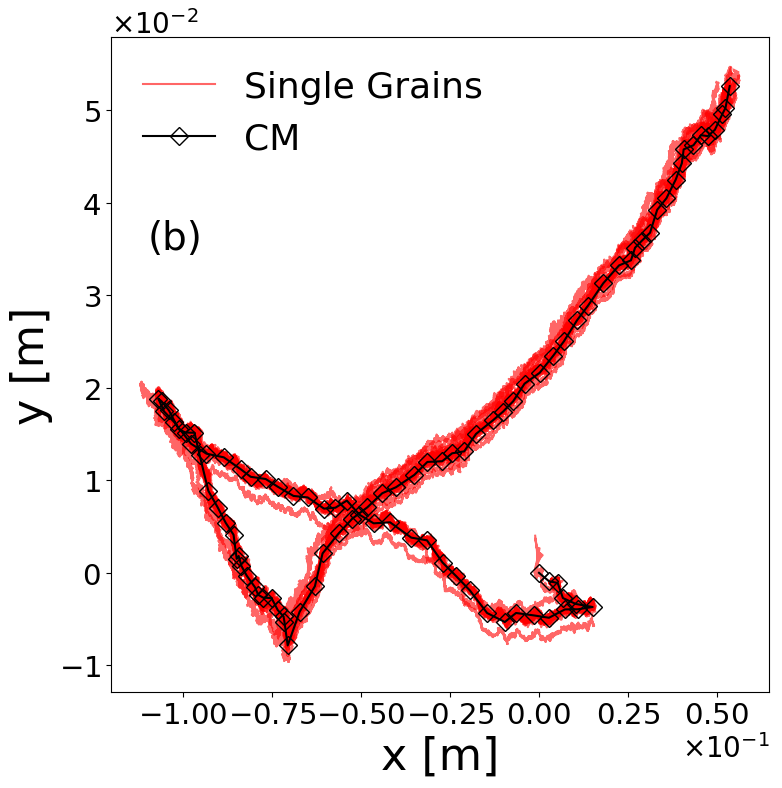}

\caption{a) Trajectories in the $xy$ plane of six grains during a time interval $\Delta t$ of a simulation in the cylindrical setup shaken at $\Gamma=39.8$. The overall dynamics can be described
as the superimposition of a slow rigid-body rotation on individual fast vibratons. The
black dot represents the dimension of a grain in scale with the graph: we see that
the amplitude of the fast fluctuations is a small with respect to the grain diameter b) Trajectories
in the $xy$ plane of a substet of 37 grains (lines) and the center of mass (symbols) in
the cubic geometry shaken at $\Gamma=10.5$.\label{fig:3Dtraj}}
\end{figure}

\section{Drfit trajectories in 3D realistic setups}
Collective drifts in realistic 3D setups are the main focus of two former publications. There, a connection between this phenomenon in the cylindrical setup and the occurrence of anomalous diffusion for a tracer has been established \cite{plati2019}. We also proposed phenomenological models to reproduce the relevant properties of the drifts \cite{Plati2020slow}. 
In the main text of this paper we then concentrate on 2D setups, while 3D goemetries are used as test-ground for the analysis of correlations between structure and dynamics. Nevertheless, in order to show explicitly how the phenomenon under study is manifested in these more realistic conditions, we report here two plots of drift trajectories in the cylindrical setup and in the cubic one (Fig. \ref{fig:3Dtraj}).

\section{Persistent Motion of Single Particles' Angular Velocity}

In this section we show that the phenomenon discovered in \cite{Moukarzel2020} (i.e. persistent rotation of vertically shaken disks around their own axis) is observed also in all the setups considered in our study. 

\begin{figure}[h!]
\centering
\includegraphics[width=0.29\textwidth]{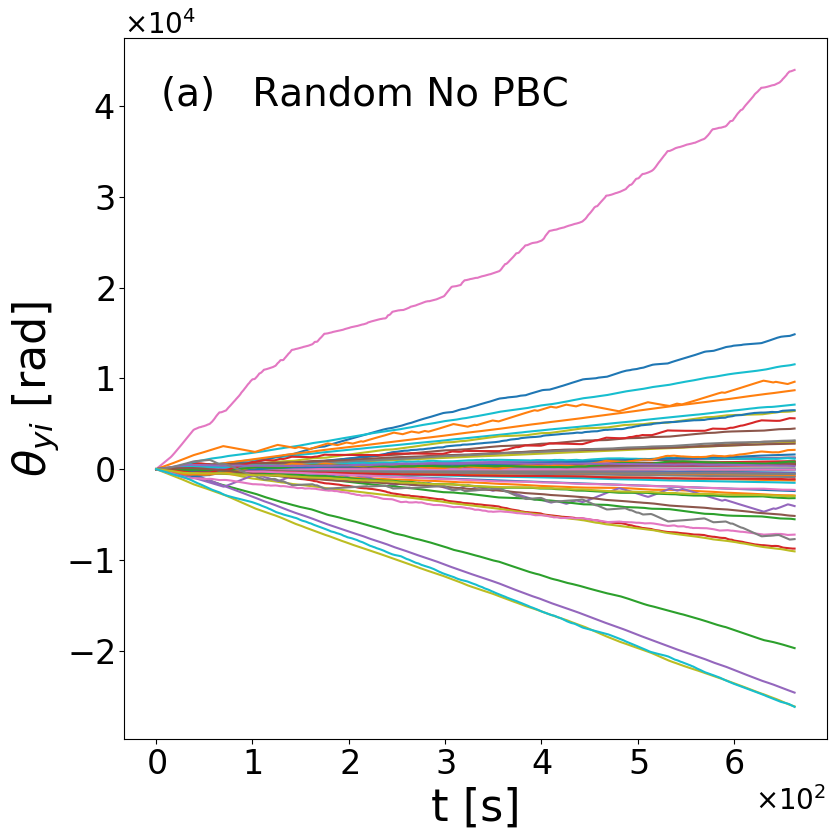}
\includegraphics[width=0.29\textwidth]{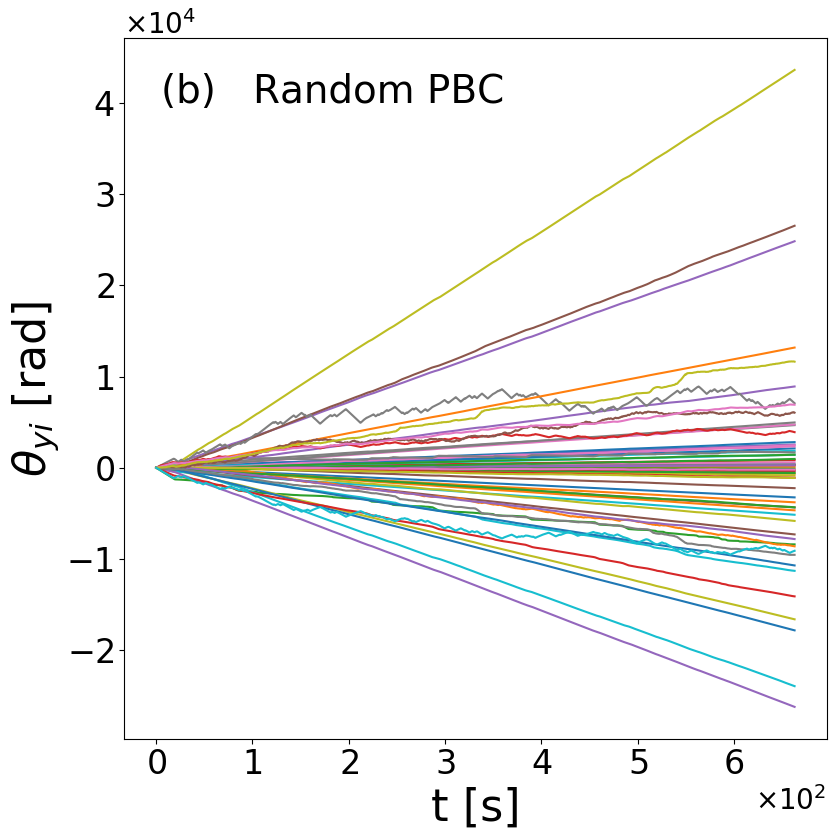}
\includegraphics[width=0.29\textwidth]{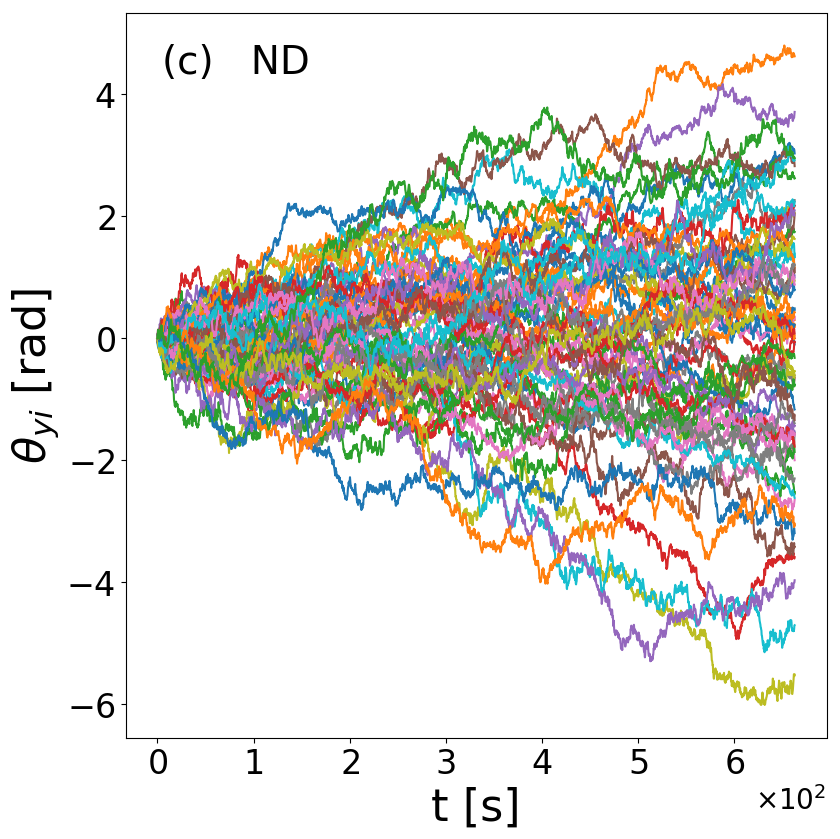}
\includegraphics[width=0.29\textwidth]{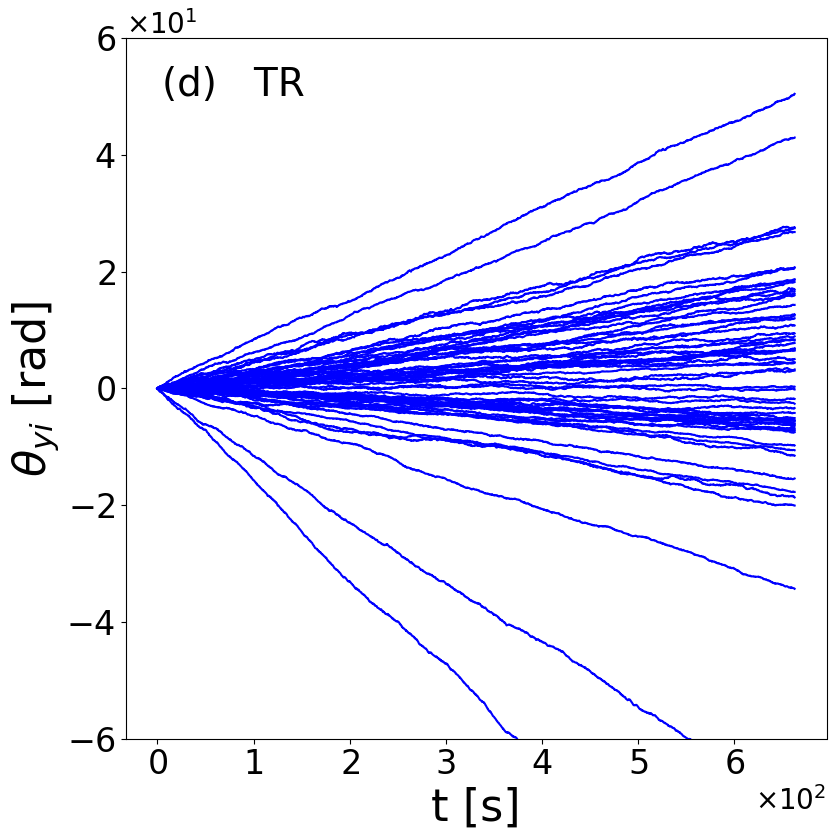}
\includegraphics[width=0.29\textwidth]{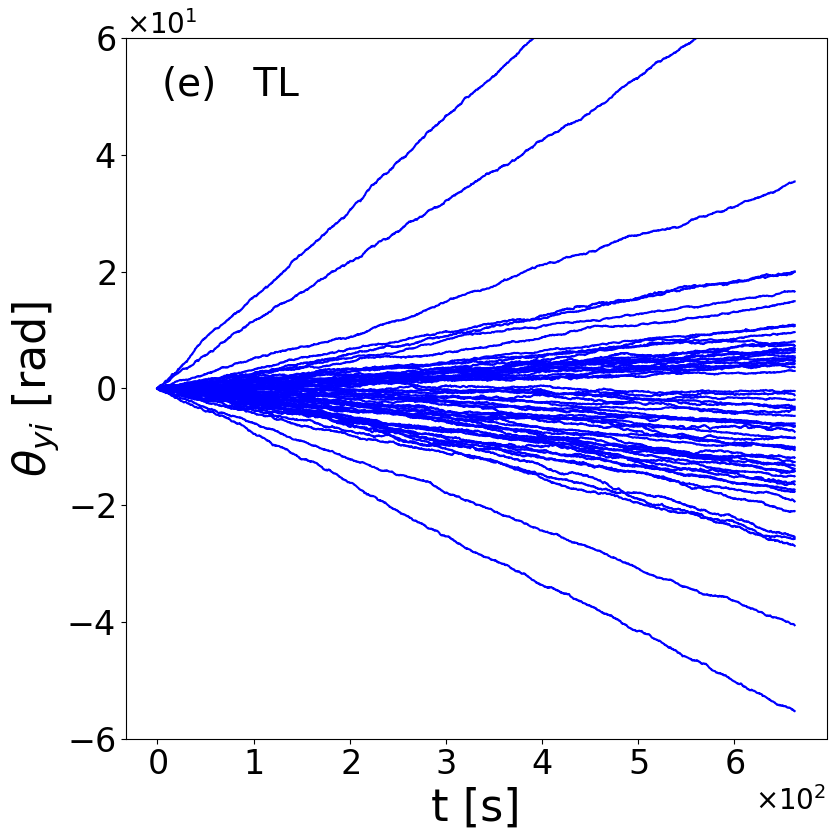}
\includegraphics[width=0.29\textwidth]{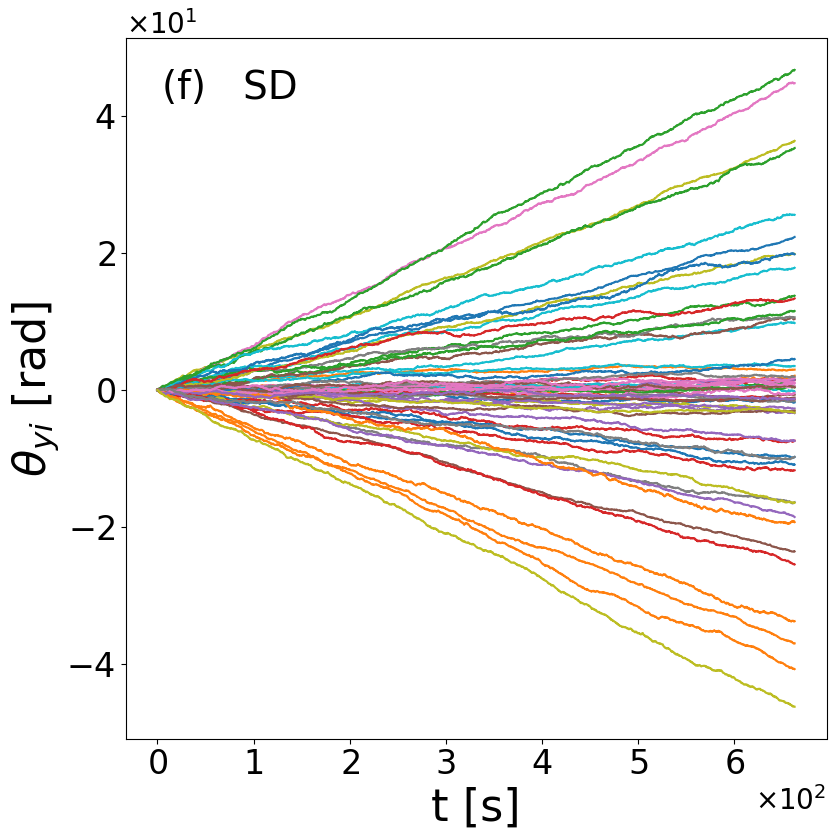}

\caption{Trajectories of the single grains absolute angles around the $y$-axis for the following cases: polydispersed packing with hard (a) and periodic (b) boundary conditions on $x$, monodispersed ordered packing with no defects (c), ordered packings with the TR (d) and RL (e) specular configurations of defects, ordered packing with the symmetric SD configuration of defects (f). All the data refer to simulations with $f=100$ and $\Gamma=0.7$ except the first panel with $\Gamma=0.85$. To see the correspondent behavior of the collective motion see Fig. 2 of the main text.\label{2Dl}}
\end{figure}

We concentrate just on the main observable related to this behavior namely the trajectories of the absolute angle $\theta_{\alpha i}(t)=\int_0^t\omega_{\alpha i}(t') dt'$ (Fig. \ref{2Dl}). Unlike disks, spherical grains can rotate around the three axis $\alpha=\{x,y,z\}$. For a direct comparison between the single particle dynamics and the collective one, we suggest to compare Fig. \ref{2Dl} with Fig. 2 of the main text.  We start by the most similar condition with respect to \cite{Moukarzel2020} considering a 2D polydispersed vertical layer confined with hard walls both in the $x$ and the $y$ direction. In Fig. \ref{2Dl}a we plot $\theta_{yi}(t)$ for all the 60 grains in the system. We can see that the single particle rotation around the $y$ axis (i.e. the equivalent that we would have had with disks) shows a persistent motion for almost all the grains in the system. Here, the presence of hard walls in the $x$ direction prevents the occurrence of the collective drift. Nevertheless, in Fig. \ref{2Dl}b, we report the same analysis performed in the equivalent setup with PBC on $x$ that allow the formation of the global translation: here is clear that the single particle persistent motion and the collective one can be observed at the same time. Panel c of the same figure shows the results obtained for the ordered monodispersed system with no defects. We see that there is not any evident long memory effect in the single particle rotational dynamics consistently with the lack of a collective translational drift.  Figs. \ref{2Dl}d-e refer to two ordered packings with specular defects and panel f to the system with a configuration of defects symmetric w.r.t. $z$. Also in all these cases the persistent motion of the grains rotation is observed but we note two additional remarkable facts: i) the distribution of the trajectory verses for TR and TL is approximately specular ii) the SD case shows long memory effects just in the single particle dynamics (we recall that no net collective drift is observed for SD). This last behavior is consistent with the claim of \cite{Moukarzel2020} that the rotational transport  is triggered by local symmetry breakings of the granular packing. The SD configuration is indeed globally symmetric but creates an asymmetric cage of nearest neighbours for many bulk particles. Regarding the mirroring of the trajectories going from TR $\to$ TL, it suggests an intriguing relation between the distribution of the $\omega_{yi}$s, the global asymmetry of the packing and consequently the persistent translational motion. In the section below we sketch some ideas on that. 
%Deepening this relation represents a promising perspective that we reserve for future works.

\begin{figure}[h!]
\centering
\includegraphics[width=0.29\textwidth]{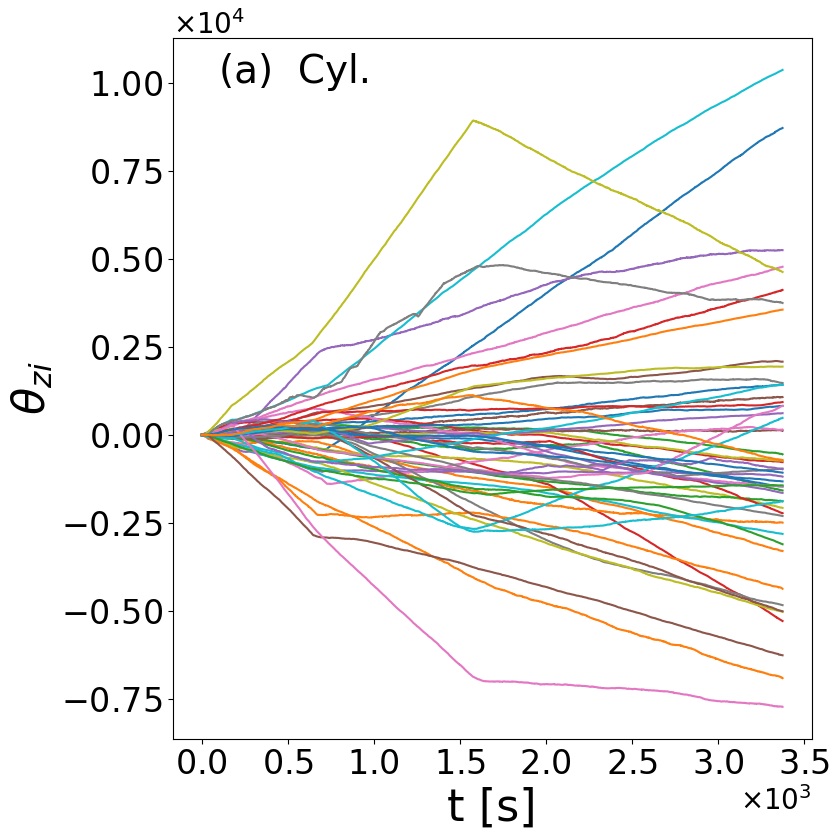}
\includegraphics[width=0.29\textwidth]{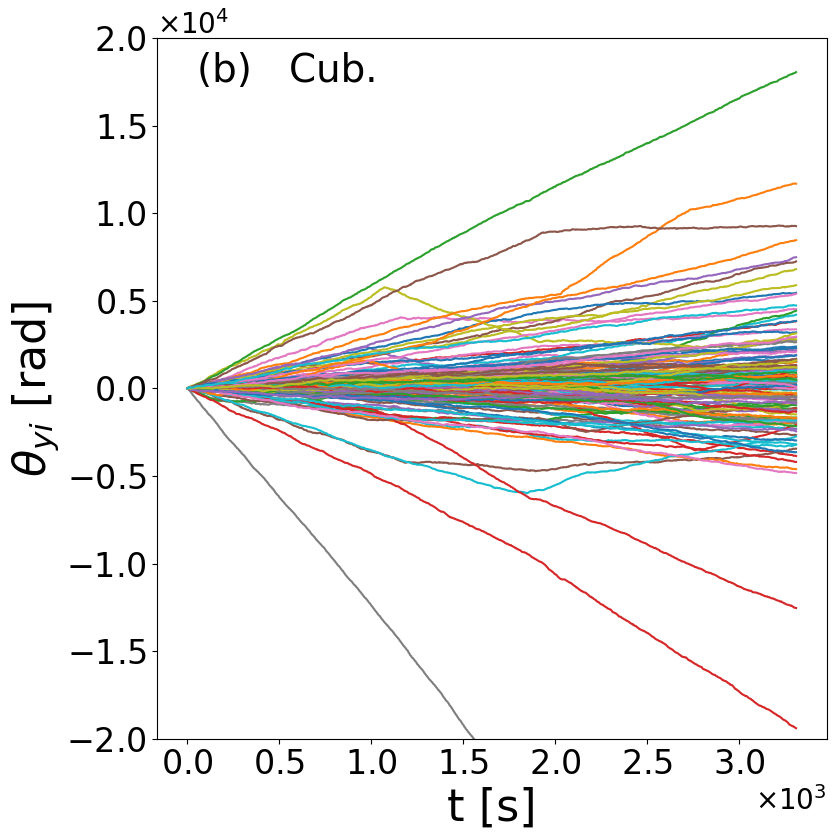}
\includegraphics[width=0.29\textwidth]{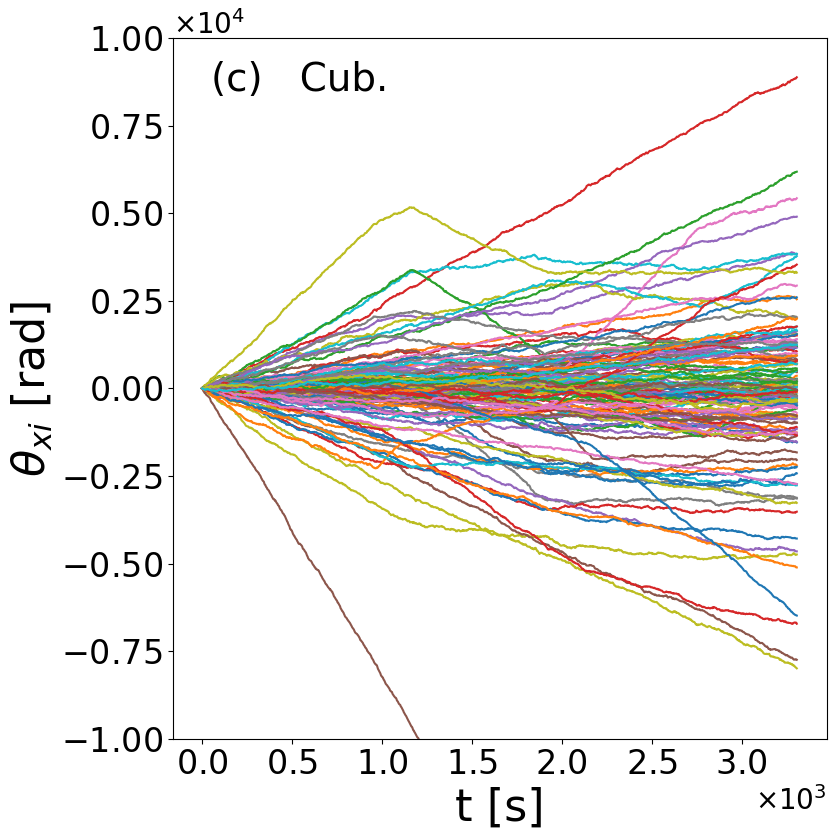}
\caption{Trajectories of the single grains absolute angles for the cylindrical setup with $\Gamma=19.5$ (a) and the cubic one with $\Gamma=10.5$ (b-c). In both cases $f=200$.\label{CylCub}}
\end{figure}

The trajectories of $\theta_{\alpha i}(t)$ for a subset of grains in the 3D setups are reported in Fig. \ref{CylCub}. We note that persistent single grains rotations are present in all the cases shown. In the cylindrical setup (panel a and b) we concentrate on $\theta_{z i}(t)$ instead of $\theta_{y i}(t)$ because, in this geometry, despite the vertical vibrations, the equivalent hard wall that is  parallel to the free direction is the lateral one. Regarding the 3D cubic setup where bot the $x$ and the $y$ directions are free we observe persistent dynamics for both  $\theta_{y i}(t)$ and $\theta_{x i}(t)$.

%single particle $z$-rotation and the revolution of the grains around the central axis of the cylinder account for the two components of the total angular momentum of the system $M=\sum_i m_i \left({\mathbf r}_i(t) \times {\mathbf v}_i(t)\right)_z + \frac{2}{5}m_iR_i^2 \omega^z_i$. 

\begin{figure}[h!]
\centering
\includegraphics[width=0.35\textwidth]{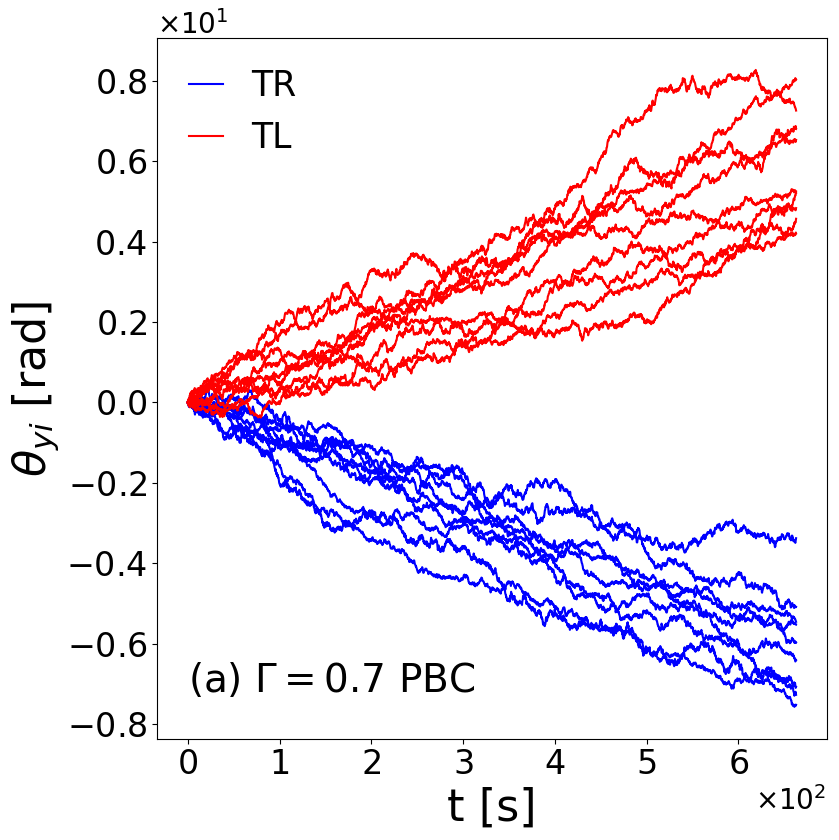}
\includegraphics[width=0.35\textwidth]{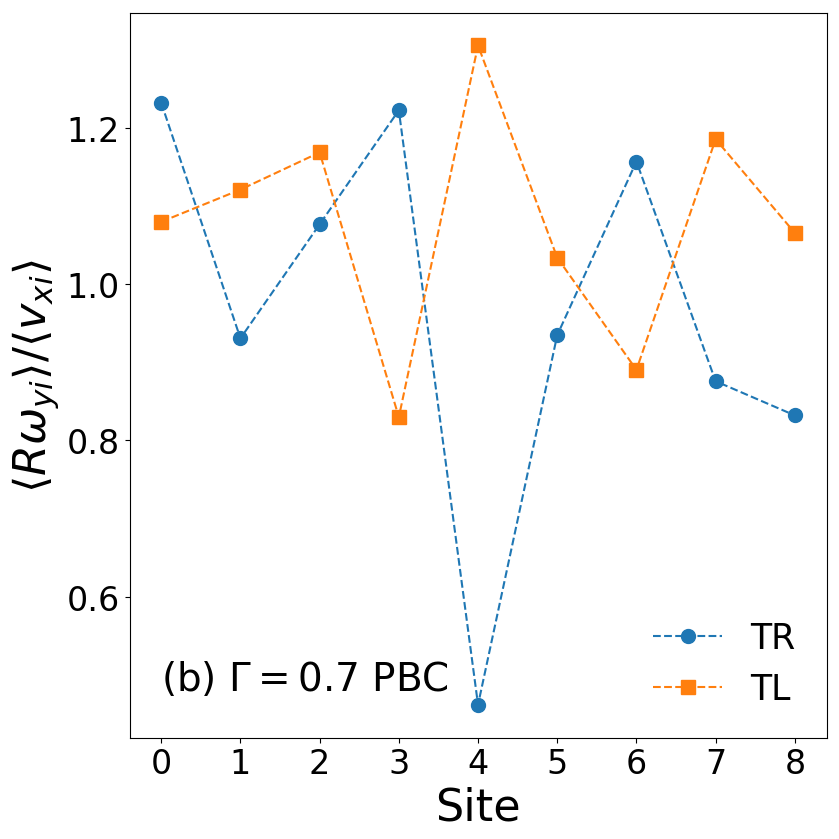}
\includegraphics[width=0.35\textwidth]{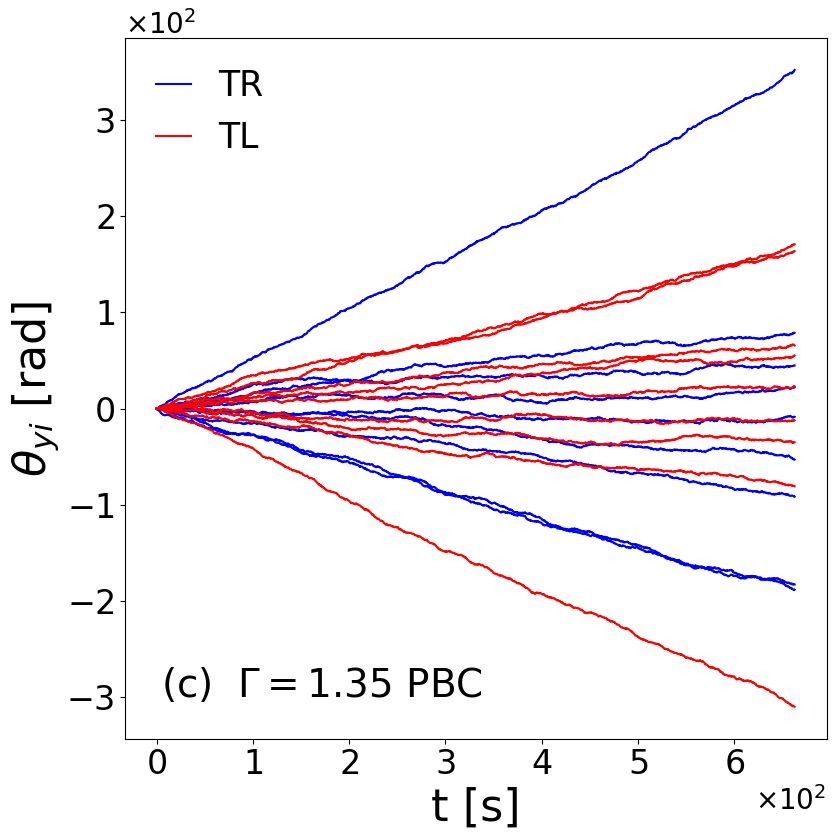}
\includegraphics[width=0.35\textwidth]{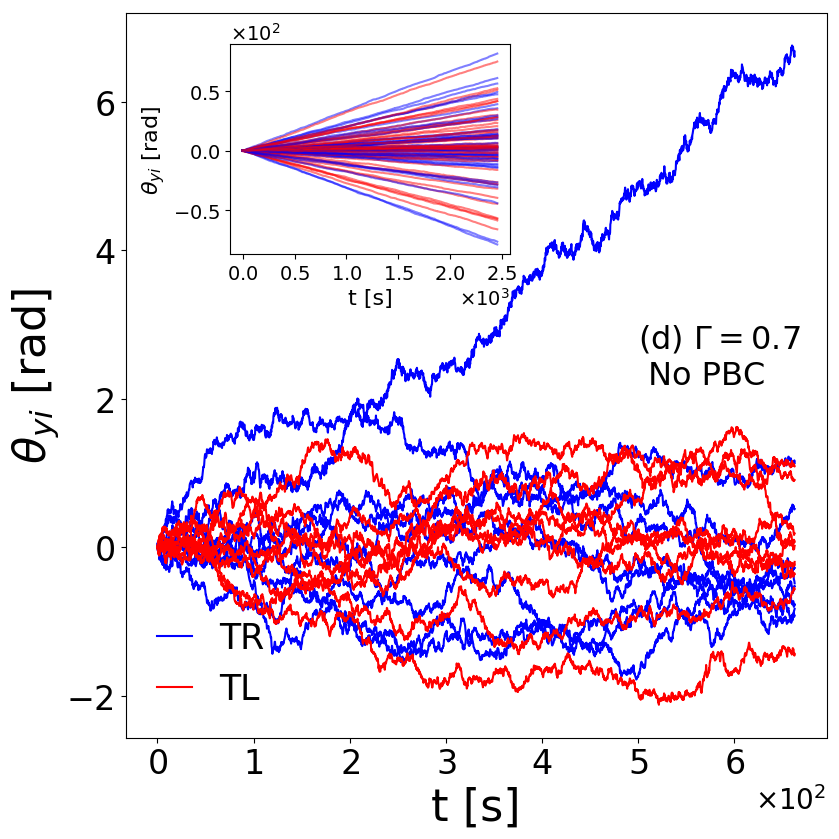}

\caption{Trajectories of the single grains absolute angle in the bottom layer for TR/TL configuration with PBC at $\Gamma=0.7$ (a), $\Gamma=1.25$ (c) and $\Gamma=0.7$ without PBC (d). In this last panel, the inset contains all the trajectories in order to show that the absence of persistent rotations regards mainly the bottom layer. In panel b, in order to check the pure rolling condition, we show the ratio of the averaged velocity at the contact point and the averaged translational one of the grains center for both TL and TR. All the data refer to simulations with $f=100$ \label{Connect}}
\end{figure}
\subsection{Connection with global translation}
The mirroring of the $\omega_{yi}$ distribution going from TR to TL hinted by Fig. \ref{2Dl}d-e suggests to study the single particle angular velocity of the bottom particles. Indeed, the total tangential force exerted by the bottom plate to the system depends on these degrees of freedom too; so they can be very important in order to understand the origin of the global dynamical asymmetry discussed in the paper (see Fig. 3e-h of the main text). 
In Fig. \ref{Connect}a we show the $\theta_{yi}$s of the bottom particles at $\Gamma=0.7$ for the two mirrored configurations. We see that, referring to the orientation of the y-axis of Fig. 1 in the main text, all the bottom particles of TR and TL rotate respectively in the negative (anti-clockwise) and positive (clockwise) verse. Remarkably, these verses correspond to the ones that, in case of pure rolling, would involve a translation with the same sign of the collective motion observed for the two configurations (i.e. negative for TR and positive for TL). From this analysis one could be tempted to conclude that the collective translation is simply the consequence of the coordinated rolling of the bottom particles acting as the "wheels" of the packing. Nevertheless, the real situation is quite more complex. First of all we can see from Fig. \ref{Connect}b, where the ratio between $R\omega_{yi}$ and $v_{xi}$ of the bottom particles is reported, that the condition of pure rolling ($\langle R\omega_{yi} \rangle/\langle v_{xi}\rangle\simeq1$) is only partially satisfied signalling that also sliding is occurring.  Moreover, going toward higher $\Gamma$ we don't observe anymore such a correspondence between the rolling of the bottom particles and the collective translation. Indeed, in Fig. \ref{Connect}c we can see that the bottom $\theta_{yi}$s for $\Gamma=1.25$ rotate in both the verses within the same configuration. We verified that in this case the ratio $\langle R\omega_{yi} \rangle/\langle v_{xi}\rangle$ is even more different from the unity (not shown). It is worth noting that also in \cite{Moukarzel2020} two different regimes of rotations at low and high $\Gamma$ were found. As a final analysis we verified that, in simulations of TR and TL \emph{without} periodic boundary conditions along the x-axis, the coordinated rolling of the bottom particles disappears and their motion is not persistent (Fig. \ref{Connect}d). However, the long memory effect of the rotations is very strong on the upper layers (see the inset).

After all these remarks, we can say that the slow collective translations and the persistent single particle rotations influence each other. But neither of these phenomena can be explained as the simple consequence of the other. In all the cases in which the collective motion is observed (i.e. when a free direction of motion is present) the persistent rotations are observed too. When the global translation is hindered by hard boundaries we still have steady rotations but with different properties with respect to the ones observed in presence PBC. As an example, we have seen that the rotational dynamics of the bottom particles for TR and TL qualitatively changes according to the presence or absence of PBC (compare panel a and d of Fig. \ref{Connect}).

To conclude, we point out that our numerical study of long memory effects in the single particles rotational degrees of freedom for spherical grains in both 2D and 3D geometries represents a considerable extension of the experimental of work reported in \cite{Moukarzel2020}. We have provided many evidences about the robustness of this phenomenon and suggested some interesting connections with the collective motion on which the main text is focused on. 
Deepening the mechanisms underlying this interplay of memory effects, its origin rooted in the structure and the resulting asymmetric interaction with the bottom plate represents a promising perspective that we reserve for future works.

\section{A stick-slip ratchet model}

Here we present the stick-slip ratchet model mentioned in the main text. Its aim is to further support the interpretation of the phenomenon under study as a ratchet effect. %We consider two variables $x$ and $x_u$ that  describe respectively the average position of the bottom layer and of the one just above it. Inspired by the numerical
%Given our observations, we wish to
%interpret our results within the framework of Brownian ratchets, proposing a variation of the well known periodically rocked ratchet \cite{Hanggi94}. As it happens for intrinsic ratchets \cite{Eichhorn2009,Kumar2008}, in our case the CM does
%not move in an external periodic potential. Nevertheless,
%it can be trapped, for small displacements, because of local
%deformations of the grains in contact with the plate:  we then replace
%the asymmetric periodic potential with a confining but slipping one. The presence
%of a stick-slip dynamics between the layers of the granular packing
%is  clearly present in  our simulations, see
%Fig. \ref{fig:Fig4}a.  
Our model describes the horizontal dynamics of
the granular medium by reducing it to two coarse-grained variables $x$
and $x_u$, 
identifying the average position of the lower and the upper region respectively. The dynamics of these coupled variables is defined in such a way to mimic the behaviour of two adjacent layers of the packing (see the main text). Then, we find reasonable to introduce in this model four crucial ingredients i.e. frictional forces, sinusoidal driving, spatial asymmetry and stick-slips. 
%the coupling between
%the two regions is non-linear and asymmetric. The mobility of the
%lower region is zero below a given tangential stress threshold, but
%jumps to a finite value above the threshold. 
Considering for simplicity an overdamped dynamics, we write down: 
%and For simplicity vibration and noise act
%only on the mobile region, i.e. on $x$. In summary we have:
\begin{subequations}\label{eq:modeltot}
\begin{align}
\dot{x}= -\frac{\mathcal{K}(\Delta x)}{\gamma}\Delta x + A_x\cos(2\pi f t) + \sqrt{2D}\eta(t) \label{eq:modela} \\
%\dot{x}_u=F_{\text{slip}} (\Delta x,s) \label{eq:modelb} \\
\dot{x}_u=s\frac{\mathcal{K}(\Delta x)}{\gamma_u}\Delta x \label{eq:modelb}
\end{align}
\end{subequations}
where $\Delta x=x-x_u$, $\eta(t)$ is a Gaussian white noise with
zero mean and unitary variance: $\langle
\eta(t)\eta(t')\rangle=\delta(t-t’)$, and $s=0,1$ denotes the state of the system (stick $s=0$, slip $s=1$). The reminiscence of viscous friction in the underdamped dynamics is given by the mobility of the two variables $\gamma^{-1}$ and $\gamma_u^{-1}$. Regarding $\mathcal{K}(\Delta x)=\left[1+\text{sign}(\Delta x)\epsilon\right]k
$, it represents a non-isotropic stiffness whose degree of asymmetry
is tuned by the adimensional parameter $-1<\epsilon <1$. From Eq. \eqref{eq:modeltot}a we see that the external driving (that directly acts just on the bottom layer) is modelled as a noise superimposed to a sinusoidal. Indeed, we find reasonable to assume that the effective force exerted in the horizontal direction keeps the same periodicity $1/f$ of the vertically vibrated plate but with a different amplitude $A_x$. The presence of noise accounts from the randomness introduced by the vertical $\rightarrow$ horizontal conversion of the excitation (it is mediated by the non linearities of the plate-grain and grain-grain contact forces). We also point out that in dense vibrofluidized granular systems, the power spectral density of orthogonal variables with respect the external forcing often exhibits a peak at the driving characteristic frequencies (Fig. \ref{fig:SystemSize}c) (see also experimental spectra in \cite{Scalliet2015}). The non-isotropic stiffness $\mathcal{K}(\Delta x)$ is the simplest way to introduce a spatial symmetry breaking in our model without directly defining a biased external force. It is further justified by a preliminary numerical analysis that suggests an effective asymmetric force between layers also in the DEM simulations (Fig. \ref{fig:Fig4}a).  

In order to implement stick-slips we use the following prescription. We define the high-strain region (HS) to be $\Delta x>\frac{\gamma F^*}{k(1+\epsilon)} \cup \Delta x< \frac{-\gamma F^*}{k(1-\epsilon)}$ and the low-strain (LS) region to be $|\Delta x|< \frac{\gamma F^*}{M k}$ with $M>\max(1+\epsilon,1-\epsilon)$, while the rest of possible values of $\Delta x$ are in the medium-strain region (MS). The state variable $s$ goes into $s=1$ when $\Delta x$ goes into HS, while it goes into $s=0$ when $\Delta x$ enters  LS. In  MS,  $s$ does not change, it depends upon its past (all simulations start in MS with $s=0$). If $s=0$ the upper variable does not move, while by tuning $\gamma_u^{-1} \gg \gamma^{-1}$ we have that as $s$ is switched to 1 a very fast slip of $x_u$ towards $x$ occurs. Looking at the behavior of two adjacent layers in a granular packing (Fig. \ref{fig:Fig4}b), we realize that the stick-slips of our model are a crude but reasonable approximation of the one observed in the DEM simulations. 

Eqs. \eqref{eq:modeltot} have been numerically integrated \cite{Wilkie2004} and the stick-slip behavior of $|\Delta x|$ is compared with the one of two adjacent DEM-simulated layers in Fig. \ref{fig:Fig4}c, while  Fig. \ref{fig:Fig4}d shows a trajectory of $x,x_u$ in the deterministic case ($D=0$). We can clearly observe a drift+oscillation dynamics of both variables.
\begin{figure}
\centering
\includegraphics[width=0.32\columnwidth,clip=true]{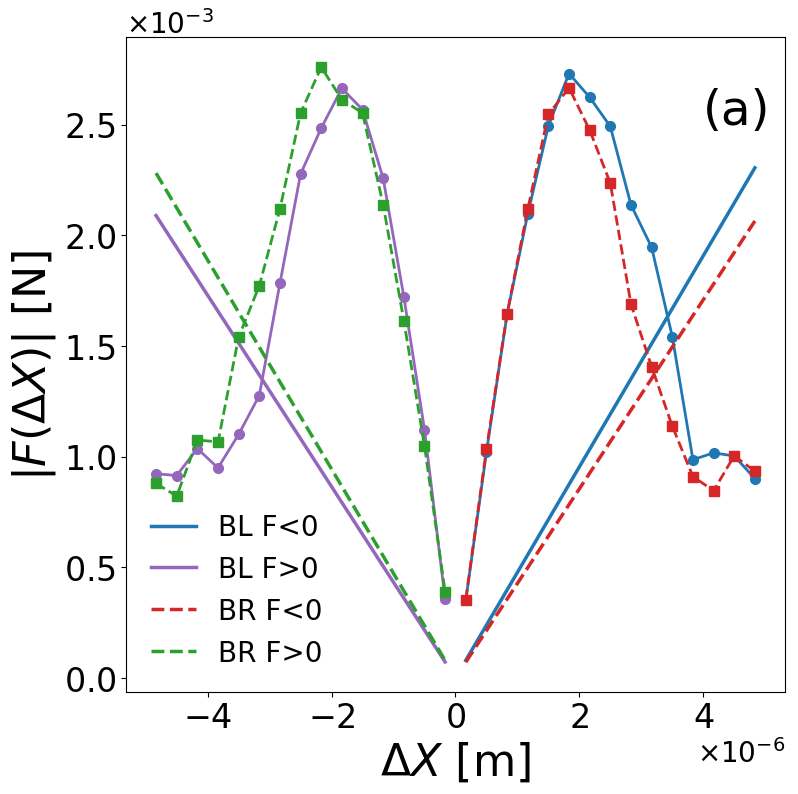}
\includegraphics[width=0.32\columnwidth,clip=true]{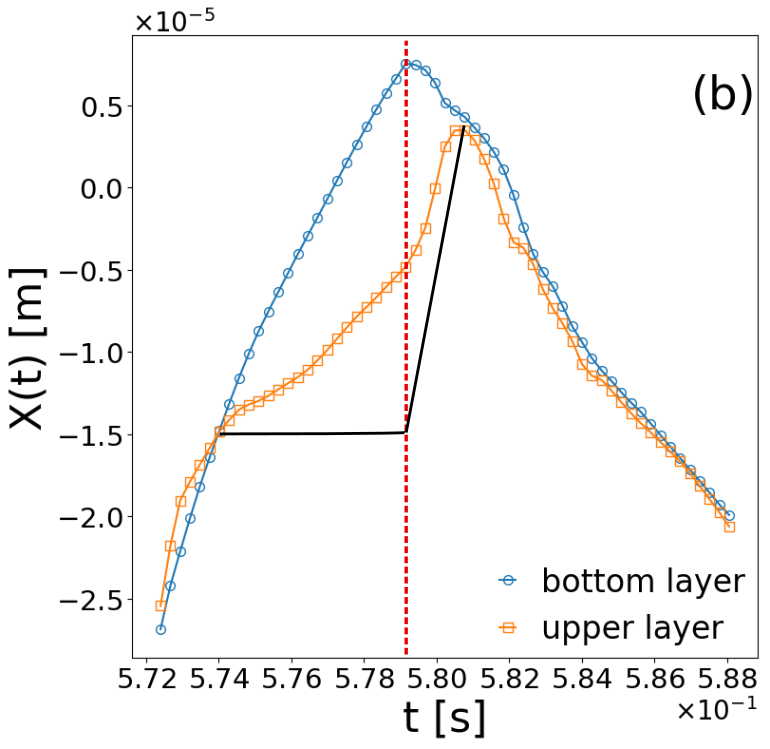}
\includegraphics[width=0.32\columnwidth,clip=true]{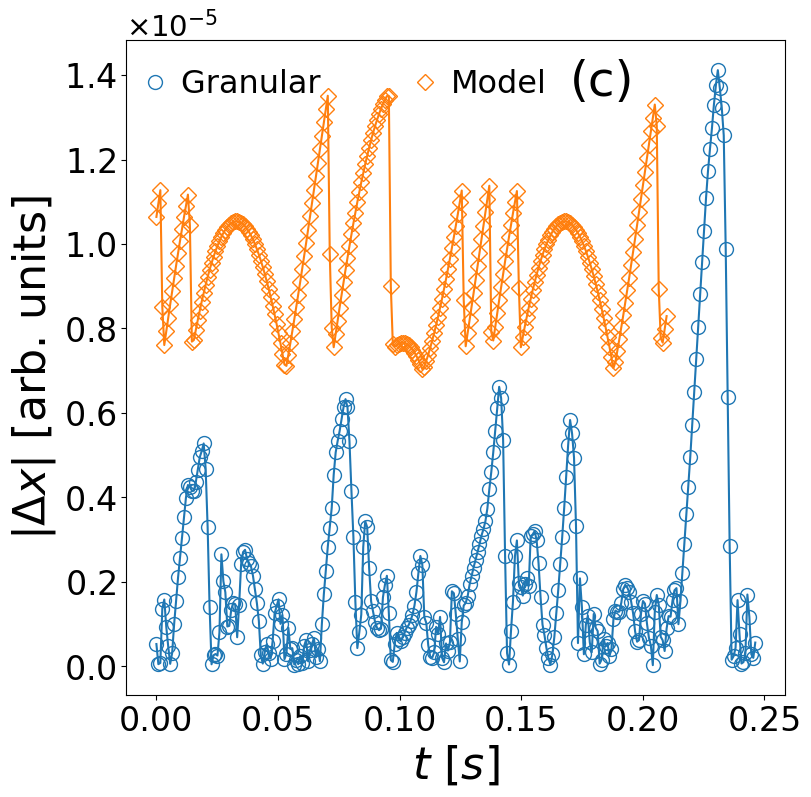}
\includegraphics[width=0.32\columnwidth,clip=true]{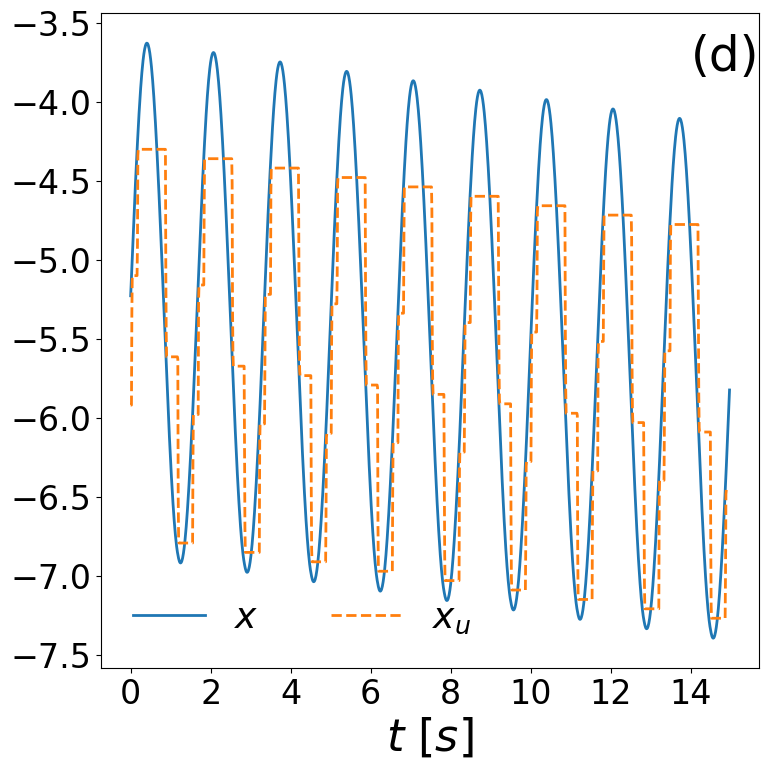}
\includegraphics[width=0.64\columnwidth,clip=true]{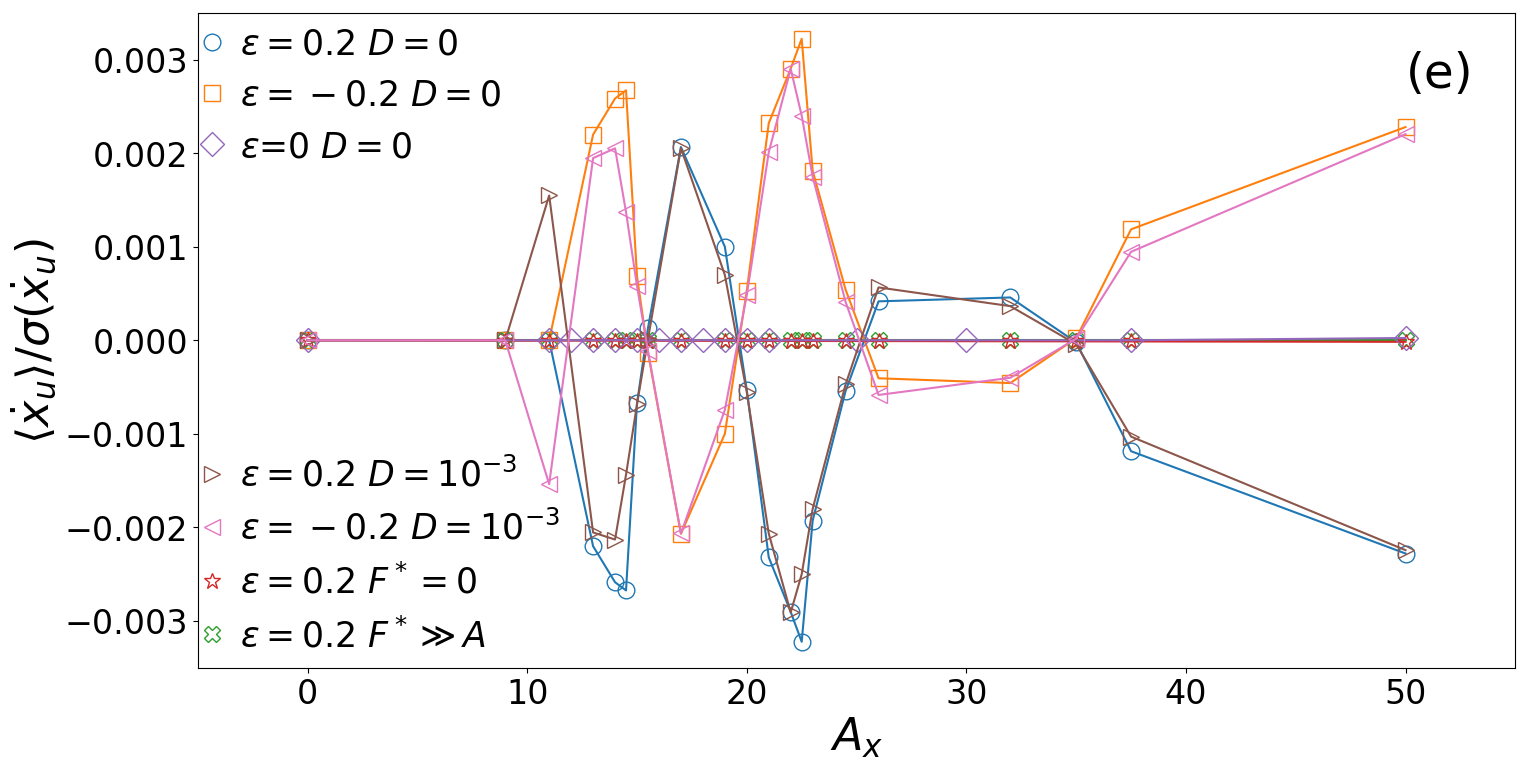}
\caption{a) Evidence of an effective asymmetric force between two adjacent layers of a granular packing in DEM simulations with $\Gamma=1.25$. Points are obtained from a scatter plot with bins, lines are linear interpolation of the data cloud. We note that the asymmetry is correctly inverted if the defects are reflected with respect to the $z$-axis.  b) Zoom of the average position of the bottom layer and the one just above it as a function of time during a stick-slip. We note that the velocity of the upper layer undergoes a sudden increase once a sufficient displacement between the two layers is reached. Black lines gives an idea of how the stick-slips implemented by our model would behave in this condition.  c) Stick-slip dynamics for the difference of the mean $x$-coordinates of two adjacent horizontal layers in a granular simulation and in the relative motion $\Delta x$ from the model. For these last two panels we considered a BL configuration with $\Gamma=1.25$. d) Typical trajectories of the model with drift and oscillations.  e) Signal to noise ratio of $\dot{x}_u$ as a function of $A_x$. For $D=10^{-3}$ the typical shape of the curve is close to the deterministic case. Values of the fixed parameters: $k=1$ s$^{-2}$, $\gamma=1$ s, $f=2$ s$^{-1}$, $\gamma_{u}^{-1}=500$ s, $M=10$, $F^*=1$.   
  \label{fig:Fig4}}
\end{figure}
%To better visualize the typical phenomenology of this dynamics, we show in the (upper left?) inset of Fig. ? a trajectory obtained by a numerical integration of Eqs. \ref{eq:modeltot} in the deterministic case ($D=0$) where we can also see that the two variables oscillate around the same mean drift.
In Fig. \ref{fig:Fig4}e we show the signal to noise ratio of
$\dot{x}_u$ as a function of $A_x$. We note that after an amplitude threshold (related to $F^*$), the
system exhibits a net motion: by setting $F^* \gg A$ the system is always trapped with a zero mean velocity. Also in the opposite limit $F^*=0$ (no stick-slip) there is no drift. The behavior of
$\dot{x}_u$ VS $A_x$ reveals a rich phenomenology, resembling granular simulation
(Fig. 2c-d) of the main text: we have drift's sign changes as $A_x$ changes with the
same $\epsilon$ and reflection of the curves as $\epsilon
\to -\epsilon$. For $\epsilon=0$ there is no mean drift, as  for symmetric packings. We stress
that the essential phenomenology of the model in the
regime where $\gamma^{-1} \ll \gamma_u^{-1}$, $f^{-1} \ll \gamma_u^{-1}$, $A_x/f \gg \sqrt{D/f}$
is contained in the
deterministic dynamics, in contrast with the ancestor model~\cite{Hanggi94}. A detailed study of the model, however, goes
beyond the scope of this work.

To summarize, our model demonstrates that frictional forces, sinusoidal driving, spatial asymmetry and stick-slips are minimal ingredients to realize the specific kind of ratchet effect that reflects the behaviour of dense vibrated granular packings. It consists of a rectification of unbiased fluctuations whose verse is determined by the interplay of internal asymmetries and external forcing. 

\section{Size effects}
\begin{figure}
\centering 
\includegraphics[width=0.33\columnwidth,clip=true]{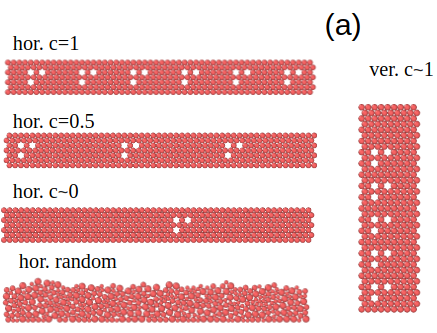}
\includegraphics[width=0.32\columnwidth,clip=true]{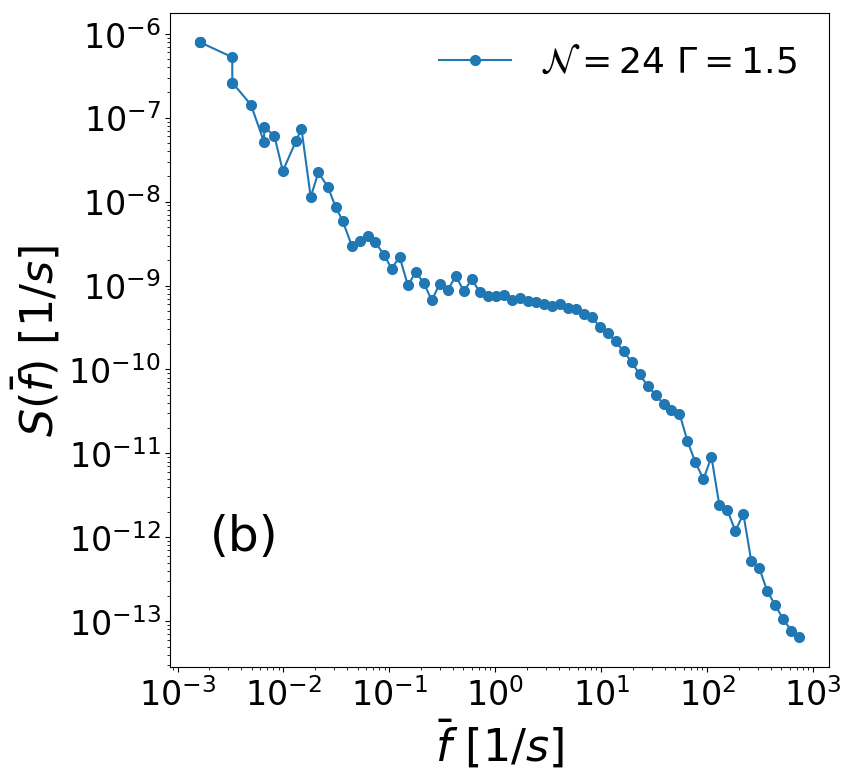}
\includegraphics[width=0.32\columnwidth,clip=true]{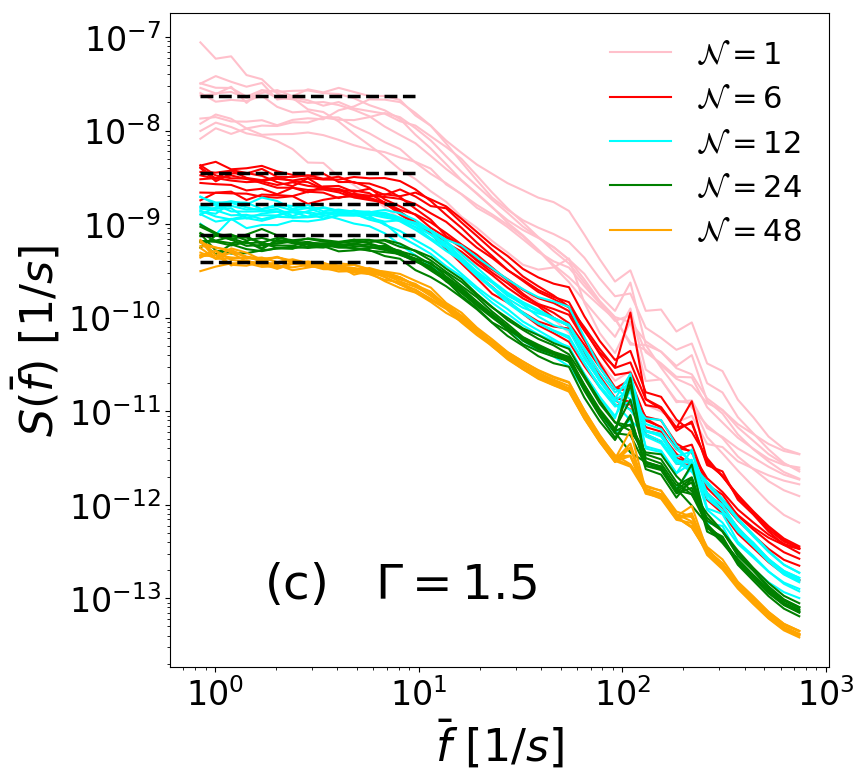}
\caption{a) Snapshots of the packings used for the study on the system's size effect. We consider cases where $\mathcal{N}=6$. In the vertical packing the highest module is without defects because we find that density fluctuations in the vertical direction tends to destroy the configurations of defects in that region of the system. b) Typical shape of the velocity power spectral density on the CM. We plot the case of a random packing with $\mathcal{N}=24$ and $\Gamma=1.5$. c) Fitted procedure for the estimate of $D$ from the plateau of the fast component of the spectra fitted with a single Lorentzian. For each $\mathcal{N}$ we considered ten independent realizations of random packings. The values of $D$ reported in the main text are calculated from the average values of the plateau whose heights are highlighted with dashed black curves in the figure. The peaks at $\bar{f}\sim 100$ Hz are related the period of the external driving.   \label{fig:SystemSize} }
\end{figure}

Here we report some supplemental informations about the study of the system's size effects. In order to better clarify the different ways in which we enlarged the size of the granular packings, we provide in Fig. \ref{fig:SystemSize}a some snapshots the simulations. We show both ordered packings with defects at different concentration $c$ and random packings. 

As mentioned in the main text, the characterization of the drift properties for large random packings requires a bit more efforts with respect what is done for ordered packings with defects. This is due to the fact that we need to estimate the asymptotic behaviour of $D$ in addition to $\langle V_x^{CM}\rangle$. 
%and also because for random packings the verse of the drift can change during the simulation. 
In order to explain the procedure used to estimate $D$/$\langle V_x^{CM}\rangle^2$, we need to point out that the global dynamics of the packings can be described as a sum of two Ornstein-Uhlenbeck processes with two well separated characteristic times \cite{Plati2020slow}. The slow component of the motion represents the global drift with very long correlations times (i.e. it decorrelates after many rearrangements that are rare events) while the fast component represents the fluctuations around the slow drift originated from the local vibrations of the grains.
As we expect from this modelization and shown in Fig. \ref{fig:SystemSize}b, the power spectral density of the CM velocity takes the form of two superimposed Lorentzian curves: 
\begin{equation}
S(\bar{f})=\lim_{t   \to \infty} \frac{1}{2\pi t}\left|\int^{t}_{0}V_x^{CM}(t')e^{2i(\pi \bar{f})t'}dt'\right|^{2}=L_1(\bar{f})+L_2(\bar{f})
\end{equation}
where $L_i(\bar{f})=\frac{D_i}{2\pi}\left[1+(2\pi \bar{f}\tau_i)^2 \right]^{-1}$. 
The low frequency decay observed in Fig. \ref{fig:SystemSize}b for $\bar{f}\lesssim 0.1$ Hz corresponds to the $1/\bar{f}^2$ decay of the slow component of the motion while for $\bar{f}\gtrsim 0.1$ Hz the spectrum is dominated by the Lorentzian representing the fast component. The 
diffusion coefficient of the fast fluctuations around the drift is then proportional to the height of the mid
frequency plateau. Indeed, for a single Lorentzian the diffusion coefficient is $D_i=2\pi L_i(0)$. Operatively, we considered only the portion of the spectrum at $\bar{f} \gtrsim 0.1$ Hz and we fitted it with a single Lorentzian $L_1(\bar{f})$ (see Fig. \ref{fig:SystemSize}c). Then, $D$ is estimated as $D=D_1=2\pi L_1(0)$. 

%Indeed, we recall that for an ordinary OU process $x(t)$ we have that $D=\lim_{t\to \infty} \langle \left [x(t)-x(0) \right]^2 \rangle/t=\lim_{f \to 0} 2\pi S(f)$.
%Figura con due o tre panel in parallelo. 1 con gli snapshots, 2 Con fitting procedure e inset (se centra di full form of VPSD)

%Regarding the typical velocities of the drift we stress that the time average of  $V_x^{CM}(t)$ can be not representative when changes of verse occur. A more  

\section{Direct visualization of the simulations}

In addition to this supplemental materials, we also provide two videos of the simulations. Video1.mp4, shows a subset of a trajectory for a random packing with PBC at high and low $\Gamma$. In the first case the system is completely fluidized, in the second one it performs a global translation on the $x$-direction. Particles rapidly vibrate in their cage and follow the collective motion on a slow time scale. Video2.mp4 contains videos for ordered packing with defects. Two configurations for two values of $\Gamma$ are considered in order to see that a sign change of the drift verse occurs reflecting the defects configuration on the $z$-axis but also varying $\Gamma$ with the same defects configuration. 
Both the videos and the snapshots in Fig. 1 of the main text has been realized with OVITO \cite{Ovito}.

\end{document}